\documentclass[12pt, preprint]{aastex}
\usepackage{amsmath}
\usepackage{graphicx}
\usepackage{natbib}

\usepackage[colorlinks=true,citecolor=blue,linktocpage=true]{hyperref}
\bibpunct{(}{)}{;}{a}{}{,} 
\usepackage[pagewise]{lineno}
\usepackage{xspace}

\usepackage{multirow}

\def\vec#1{\ensuremath{\mathbf{#1}}}

\shorttitle{Sausage modes in nonuniform loops with finite plasma $\beta$}
\shortauthors{Chen et al.}

\begin{document}

\title{FAST SAUSAGE MODES IN MAGNETIC TUBES WITH CONTINUOUS TRANSVERSE PROFILES:
       EFFECTS OF A FINITE PLASMA BETA}

\author{Shao-Xia Chen\altaffilmark{1}}
\author{Bo Li\altaffilmark{1}}
    \email{bbl@sdu.edu.cn}
\author{Ming Xiong\altaffilmark{2}}
\author{Hui Yu\altaffilmark{1}}
\and
\author{Ming-Zhe Guo\altaffilmark{1}}

\altaffiltext{1}{Shandong Provincial Key Laboratory of Optical Astronomy and Solar-Terrestrial Environment, Institute of Space Sciences, Shandong University,Weihai, 264209, China}
\altaffiltext{2}{National Space Science Center, CAS, 100190 Beijing, China}

\begin{abstract}
While standing fast sausage modes in flare loops are often invoked to interpret quasi-periodic pulsations (QPPs)
    in solar flares, it is unclear as to how they are influenced
    by the combined effects of a continuous transverse structuring and a finite internal plasma beta ($\beta_{\rm i}$).
We derive a generic dispersion relation (DR) governing linear sausage waves in straight magnetic tubes
    for which plasma pressure is not negligible and
    the density and temperature inhomogeneities of essentially arbitrary form take place in a layer of arbitrary width.
Focusing on fast modes, we find that $\beta_{\rm i}$ only weakly influences $k_{\rm c}$,
    the critical longitudinal wavenumber separating the leaky from trapped modes.
Likewise, for both trapped and leaky modes,
    the periods $P$ in units of the transverse fast time depend only weakly on $\beta_{\rm i}$,
    which is compatible with the fact that the effective wavevectors of
    fast sausage modes are largely perpendicular to the background magnetic field.
However, a weak $\beta_{\rm i}$ dependence of the damping times $\tau$ is seen only when the length-to-radius ratio
    $L/R$ is $\sim 50\%$ larger than some critical value $\pi/(k_{\rm c} R)$, which itself
    rather sensitively depends on the density contrast, profile steepness as well as
    on how the transverse structuring is described.
In the context of QPPs, we conclude that the much simpler zero-beta theory can be employed for trapped modes,
    as long as one sees the deduced internal Alfv\'en speed as actually being the fast speed.
In contrast, effects due to a finite beta in flare loops should be considered when leaky modes are exploited.
\end{abstract}

\section{INTRODUCTION}
\label{sec_intro}
Recent years have seen rapid progress
    in the field of solar magneto-seismology (SMS),
    thanks to the abundantly identified low-frequency waves and oscillations in the solar atmosphere
    \citep[for recent reviews, see e.g.,][]{2005LRSP....2....3N,2007SoPh..246....3B,2012RSPTA.370.3193D,2016GMS...216..395W,2016SSRv..200...75N}.
While originally proposed in the coronal context \citep{1970A&A.....9..159R, 1970PASJ...22..341U, 1975IGAFS..37....3Z}
    and hence named coronal seismology \citep{1984ApJ...279..857R},
    SMS has been extended to other parts of the Sun's atmosphere
    such as spicules \citep[e.g.][]{2009SSRv..149..355Z},
    prominences \citep[e.g.,][]{2012LRSP....9....2A},
    pores and sunspots \citep[e.g.,][]{2008IAUS..247..351D, 2011ApJ...729L..18M, 2014A&A...563A..12D},
    as well as various chromspheric structures \citep[e.g.,][]{2009Sci...323.1582J,2012NatCo...3E1315M}.
In addition, the ideas behind SMS are not restricted to inferring the physical parameters of localized structures,
    but also have found applications in the so-called ``global coronal seismology''~\citep{2005A&A...435.1123W,2007SoPh..246..177B}
    where various large-scale coronal waves are exploited
    to deduce the global magnetic field in the corona
    \citep[see][for recent reviews]{2014SoPh..289.3233L,2015LRSP...12....3W, 2016GMS...216..381C}.

{The modern terminology in SMS for mode classification
    largely comes from
    \citet{1983SoPh...88..179E}, where the rich variety of linear collective
    wave modes are examined for straight magnetized tubes aligned with the equilibrium magnetic field
    \citep[see also the reviews by][]{2005LRSP....2....3N,2008IAUS..247....3R}.
Kink modes correspond to case where the azimuthal wavenumber $m=1$, and are the only modes that displace the tube axis.
On the other hand, sausage modes are axisymmetric, corresponding to the case where $m=0$.
Both kink  and sausage modes
}
    are important for applications in SMS, even though kink ones
    seem to have attracted more attention
    \citep[e.g.,][]{1999Sci...285..862N,1999ApJ...520..880A,2009ApJ...697.1384T,2013SoPh..284..559K, 2013A&A...560A.107A, 2015A&A...583A.136A}.
As a matter of fact, recent observations indicated that sausage modes abound in the solar atmosphere as well.
Sausage waves were found to be ubiquitous together with kink waves in the chromosphere~\citep{2012NatCo...3E1315M},
   and their signatures have been found in pores and sunspots
   \citep[e.g.,]{2011ApJ...729L..18M, 2014A&A...563A..12D,2015A&A...579A..73M, 2015ApJ...806..132G,2016ApJ...817...44F}.
On top of that, fast sausage modes in flare loops have long been
    suggested to account for quasi-periodic pulsations (QPPs) with periods of the order of seconds
    in the lightcurves of solar flares
    (\citeauthor{1970A&A.....9..159R}~\citeyear{1970A&A.....9..159R};
    \citeauthor{1975IGAFS..37....3Z}~\citeyear{1975IGAFS..37....3Z};
    and reviews by \citeauthor{1987SoPh..111..113A}~\citeyear{1987SoPh..111..113A};
    \citeauthor{2009SSRv..149..119N}~\citeyear{2009SSRv..149..119N}).
While QPPs were primarily examined with spatially unresolved observations prior to 2000~\citep{2004ApJ...600..458A},
    they can now be readily measured using imaging instruments such as
    the Nobeyama Radioheliograph \citep[NoRH, e.g.,][]{2001ApJ...562L.103A,  2003A&A...412L...7N, 2005A&A...439..727M, 2013SoPh..284..559K},
    the Atmospheric Imaging Assembly on board the Solar Dynamics Observatory \citep[SDO/AIA,][]{2012ApJ...755..113S},
    and more recently with the Interface Region Imaging Spectrograph \citep[IRIS,][]{2016ApJ...823L..16T}.

To facilitate proper seismological applications, a substantial number of theoretical and numerical studies
    have been conducted to examine sausage waves collectively supported by magnetized tubes
    \citep[e.g.,][]{1978SoPh...58..165M, 1982SoPh...75....3S, 1983SoPh...88..179E,1986SoPh..103..277C,2007AstL...33..706K,
    2012ApJ...761..134N, 2014A&A...572A..60L,2014ApJ...781...92V, 2015ApJ...812...22C,2016SoPh..291..877G}.
Let $\rho$ denote plasma density, and
     let $v_{\mathrm A}$ and $c_{\mathrm s}$ denote the Alfv\'en and sound speeds, respectively.
Furthermore, let subscript ${\mathrm i}$ (${\mathrm e}$) denote the parameters inside (outside) a tube.
In an environment such as the corona where the ordering $v_\mathrm{Ae} > v_\mathrm{Ai} >c_\mathrm{si} >c_\mathrm{se}$ holds,
    two regimes of fast sausage modes are known to exist, depending on the relative magnitude of longitudinal wavenumber $k$
    with respect to a critical value $k_\mathrm{c}$.
The trapped regime results when $k>k_\mathrm{c}$ whereby sausage modes are well confined,
    whereas the leaky regime arises when $k<k_\mathrm{c}$ whereby sausage modes experience apparent damping
    because oscillating tubes continuously emit fast waves into their surroundings
    \citep[e.g.,][]{1982SoPh...75....3S, 1986SoPh..103..277C}.
When $k$ is sufficiently small, neither the periods ($P$) nor the damping times ($\tau$)
    of leaky fast sausage waves depend on $k$ any more
    \citep[e.g.,][]{2007AstL...33..706K, 2012ApJ...761..134N, 2014ApJ...781...92V, 2015SoPh..290.2231C, 2015ApJ...812...22C}.
This then makes it easier to invert the measured values of $P$ and $\tau$
    to deduce information on the magnetic and plasma structuring for the tubes hosting sausage modes,
    provided that these tubes can be supposed to be sufficiently thin.
Take the simplest case where this structuring is in the step-function (top-hat) form for instance.
It turns out that $\tau/P \approx (\rho_{\rm i}/\rho_{\rm e})/\pi^2$
    and $P \approx 2.62 R/\sqrt{c_{\mathrm{si}}^2+v_{\mathrm{Ai}}^2}$ when $k \rightarrow 0$ and $\rho_\mathrm{i}/\rho_\mathrm{e} \gg 1$,
    where $R$ represents the tube radius
    \citep{2007AstL...33..706K}.
With $P$ and $\tau$ known from QPP measurements, it is then possible to deduce both $\rho_\mathrm{i}/\rho_\mathrm{e}$
    and $v_{\mathrm{Ai}}$, the latter carrying the information on the magnetic field strength in the key region
    where flare energy is released.

For mathematical simplicity, theoretical studies of sausage modes in magnetic tubes tended to
    invoke one or both of the following two assumptions:
    one is the cold (zero-$\beta$) MHD limit where thermal pressure is neglected,
    the other is that the plasma and magnetic parameters are transversally structured in a top-hat fashion.
While lifting the second assumption by adopting a continuous transverse profile,
    the analytical studies by \citet{2014A&A...572A..60L, 2015ApJ...810...87L, 2015ApJ...812...22C}
    nonetheless worked in the zero-$\beta$ limit.
{We note by passing that the effects of continuous transverse profiles on sausage modes have also been examined
    for pressureless slabs~\citep[e.g.,][]{2015ApJ...801...23L, 2015ApJ...814...60Y}.
When}
    addressing the effects of a finite plasma $\beta$,
    the analytical works by \citet{1983SoPh...88..179E,  2007AstL...33..706K}
    adopted top-hat profiles to model the transverse distributions of the magnetic
    and plasma parameters.
Evidently, physical parameters are more likely to be continuously structured transverse to tubes.
On the other hand, the plasma $\beta$ is not necessarily small but may reach a value of order unity
    in hot and dense loops in both active regions \citep{2007ApJ...656..598W}
    and flares \citep[e.g.,][]{2005A&A...439..727M}.
There is therefore an evident need to develop
    a theoretical description incorporating both effects of a finite $\beta$
    and a continuous transverse profile.
The aim of the present manuscript is to offer such a description,
    which can be seen as a natural extension
    to our previous work
    \citep[][hereafter paper I]{2015ApJ...812...22C} where we adopted the framework of cold MHD
    to formulate an analytical dispersion relation (DR) for sausage waves in tubes with
    essentially arbitrary transverse density profiles.
{We will focus on linear sausage waves in straight magnetized tubes aligned with the equilibrium magnetic field.
In addition, we will work in the framework of ideal MHD, meaning that the damping of sausage waves is not due to dissipative processes
   but a result of lateral leakage.
}

Before proceeding, we note that
    sophisticated numerical simulations can certainly address both of the above-mentioned effects simultaneously.
However, so far the only dedicated numerical studies on sausage waves
    in a cylindrical geometry \citep{2004A&A...422.1067S,2015ApJ...814..135S}
    were primarily interested in examining the temporal signatures of impulsively generated waves
    rather than providing
    a detailed investigation on their dispersive properties.
In fact, developing an analytical DR is important not only in its own right,
    but also helps better understand these numerical results.
The reason is that, the temporal and wavelet signatures of impulsively generated sausage waves
    depend critically on
    the frequency dependence of the longitudinal group speeds of trapped modes
    \citep{1983Natur.305..688R, 1984ApJ...279..857R}.
With the DR to be developed, such a frequency dependence can be readily evaluated.
{We further note that \citet{2009A&A...503..569I} have carried out a numerical study on
    the effects of a finite plasma $\beta$ on trapped standing sausage modes in
    coronal slabs, for which the equilibrium parameters are continuously distributed in the transverse direction.
These authors found that plasma $\beta$ has only a weak influence on both the periods of the fundamental modes
    and the cutoff wavenumber that separates trapped from leaky regimes.
Our study differs from \citet{2009A&A...503..569I} in two aspects.
First, we will adopt a cylindrical geometry to examine sausage modes in magnetized tubes for which the transverse distribution
    of equilibrium parameters is rather general.
Second, an eigenmode analysis will be carried out to enable the derivation of an analytical dispersion relation of sausage modes that is valid
    in both trapped and leaky regimes.
Similar to \citet{2009A&A...503..569I}, we will also solve the time-dependent ideal MHD equations to examine sausage perturbations from an initial-value-problem perspective.
The results thus found will be used to validate our eigenmode analysis.
Despite these differences, our analysis will show that in the cylindrical geometry,
    the cutoff wavenumber also shows only a weak dependence on plasma $\beta$.
The periods and damping times of sausage modes also only weakly depend on plasma $\beta$, as long as they are measured
    in units of the time it takes for fast waves to traverse the cylinder.
}

This manuscript is organized as follows.
In Section \ref{sec_equilibrium} we present the necessary description
    for the parameters characterizing a magnetic tube.
The derivation of the DR is given in Sect.~\ref{sec_DR},
    and we then offer in Sect.~\ref{sec_para_study}
    a rather detailed examination of the effects due to a finite $\beta$.
Section~\ref{sec_conclusion} closes this manuscript with our summary
    and some concluding remarks.

\section{DESCRIPTION FOR THE EQUILIBRIUM TUBE}
\label{sec_equilibrium}

\subsection{Overall Description}
\label{sec_sub_equil_overall}
We model coronal loops as straight magnetized tubes and establish a standard
     cylindrical coordinate system $(r, \theta, z)$ where the $z$-axis is aligned with the tube.
The equilibrium magnetic field $\vec{B}$ is also in the $z$-direction.
Both the plasma parameters and magnetic field strength are assumed to be a function of $r$ only.
Let $p$ denote thermal pressure.
It then follows from the transverse force balance condition that
\begin{equation}
     p(r)+\displaystyle\frac{B^2(r)}{8\pi}= \mbox{const} \equiv \alpha .
\label{eq_force_balance}
\end{equation}
Restricting oneself to an electron-proton plasma, one finds that $p$ is related to the density $\rho$ and temperature $T$
     via
\begin{equation}
 p = \displaystyle\frac{2 k_{\rm B}}{m_p} \rho T ,
\label{eq_state}
\end{equation}
    with $k_{\rm B}$ being the Boltzmann constant and $m_p$ the
    proton mass.
In view of Eqs.~(\ref{eq_force_balance}) and (\ref{eq_state}), one can arbitrarily specify the transverse profiles
    for any two out of the three quantities $[\rho, T, B]$.
Without loss of generality, we choose to specify $\rho(r)$ and $T(r)$.

The following characteristic speeds are necessary for examining sausage waves.
To start, the adiabatic sound and Alfv\'en speeds are given by
 \begin{equation}
   c_{\rm s}^2=\displaystyle\frac{\gamma p}{\rho},
   \hspace{0.5cm} \mbox{and} \hspace{0.5cm}
   v^2_{\rm A}=\displaystyle\frac{B^2}{4\pi\rho},
   \label{eq_def_cs_va}
 \end{equation}
     where $\gamma=5/3$ is the adiabatic index.
When expressed in terms of $c_{\rm s}$ and $v_{\rm A}$, the plasma $\beta \equiv 8\pi p/B^2$ reads
\begin{equation}
   \beta = \frac{2}{\gamma}\frac{c^2_{\rm s}}{v^2_{\rm A}} .
   \label{eq_def_beta}
\end{equation}
The fast speed is then defined to be
 \begin{equation}
   v_{\rm f}^2= c_{\rm s}^2 +v^2_{\rm A},
   \label{eq_def_vf}
 \end{equation}
     which, strictly speaking, pertains to perpendicular propagation in a uniform equilibrium.
Finally, the tube speed is defined by
\begin{equation}
    c_{\rm T}^2=\displaystyle\frac{c_{\rm s}^2v^2_{\rm A}}{c_{\rm s}^2+v^2_{\rm A}} .
    \label{eq_def_ct}
\end{equation}

\subsection{Description for Transverse Profiles}
Evidently, the profiles for $\rho(r)$ and $T(r)$ are independent from each other.
To avoid our derivation becoming unnecessarily too lengthy, however,
    we assume that $\rho$ and $T$
    have the same formal dependence on $r$.
To be specific, they are described by
\begin{eqnarray}
 {\rho}(r)=\left\{
   \begin{array}{ll}
   \rho_{\rm i},    							& 0\le r \leq r_{\rm i} = R-l/2 ,\\
   \rho_{\rm tr}(r) = {\cal F}(\rho_{\rm i}, \rho_{\rm e}; r), 		& r_{\rm i} \le r \le r_{\rm e} = R+l/2, \\ \label{eq_profile_rho_overall}
   \rho_{\rm e},    							& r \ge r_{\rm e} ,
   \end{array}
   \right.
\end{eqnarray}
    and
\begin{eqnarray}
 T(r)=\left\{
   \begin{array}{ll}
   T_{\rm i},    				& 0\le r \leq r_{\rm i} , \\
   T_{\rm tr}(r) = {\cal F}(T_{\rm i}, T_{\rm e}; r),	& r_{\rm i} \le r \le r_{\rm e} ,\\ \label{eq_profile_T_overall}
   T_{\rm e},    				& r \ge r_{\rm e}.
   \end{array}
   \right.
\end{eqnarray}
In other words, the equilibrium configuration is assumed to comprise a uniform cord (denoted
    by subscript ${\rm i}$), a uniform external medium (subscript ${\rm e}$),
    and a transition layer (TL) connecting the two.
This TL is of width $l$ and centered around the mean tube radius $R$.
Furthermore, ${\cal F}(\epsilon_{\rm i}, \epsilon_{\rm e}; r)$ is some function that smoothly connects
    $\epsilon_{\rm i}$ at the cord-TL interface ($r_{\rm i}$)
    to $\epsilon_{\rm e}$ at the TL-external-medium interface ($r_{\rm e}$).

That ${\cal F}$ is smooth in the interval $[r_{\rm i}, r_{\rm e}]$ makes it possible to Taylor expand $\rho_{\rm tr}$
    and $T_{\rm tr}$ around $x\equiv r-R = 0$, resulting in
\begin{eqnarray}
  \rho_{\rm tr}(x) = \sum^\infty_{n=0}\rho_n x^n,~~~~~~~~T_{\rm tr}(x) = \sum^\infty_{n=0}T_n x^n ,
\label{eq_rhoT_expansion}
\end{eqnarray}
    where $\rho_0 = \rho|_{x=0}$, $T_0 = T|_{x=0}$ and
\begin{eqnarray}
\label{eq_rho_coef}
  \rho_n = \frac{1}{n!} \left.\frac{{\rm d}^n\rho(x)}{{\rm d}x^n}\right|_{x=0},~~~~~~~~~~~T_n = \frac{1}{n!} \left.\frac{{\rm d}^nT(x)}{{\rm d}x^n}\right|_{x=0},~~~~~~~~~~~
     \hspace{0.2cm} n\ge 1 .
\end{eqnarray}
In the TL, $c_{\rm s}^2$ can be expanded as $c_{\rm s}^2 = \sum\limits^\infty_{n=0}C_n x^n$ with
\begin{equation}
  C_n=\displaystyle\frac{2\gamma k_{\rm B}}{m_p}T_n ,
  \label{eq_coef_Cn}
\end{equation}
    which is a direct result of the definitions~(\ref{eq_state}) and (\ref{eq_def_cs_va}).
To derive the explicit form for the coefficients in the expansion $v_{\rm A}^2 = \sum\limits^\infty_{n=0}V_n x^n$,
    we start with reformulating Eq.~(\ref{eq_force_balance}) in terms of $c_{\rm s}^2$ and $v_{\rm A}^2$, arriving at
 \begin{equation}
\rho (x) v^2_{\rm A} (x)
    =2\alpha- \displaystyle\frac{2 \rho (x) c^2_{\rm s}(x)}{\gamma}~,
\label{eq_forcebalance_rhovacs}
 \end{equation}
    or equivalently,
\begin{equation*}
  \left(\sum^\infty_{n=0}\rho_n x^n\right)
  \left(\sum^\infty_{n=0}   V_n x^n\right)
= 2\alpha
  -\displaystyle\frac{2}{\gamma}
    \left(\sum^\infty_{n=0}\rho_n x^n\right)
    \left(\sum^\infty_{n=0}C_n x^n\right) .
 \end{equation*}
Manipulating the product of two series, e.g.,
    $\left(\sum\limits^\infty_{n=0}\rho_n x^n\right)\left(\sum\limits^\infty_{n=0}V_n x^n\right)
    =\sum\limits^\infty_{l=0}\sum\limits^l_{n=0}V_n \rho_{l-n} x^l$,
   and equating the coefficient of $x^l$, one finds that
\begin{equation}
  \left\{
  \begin{array}{rcl}
  V_0 &=& \displaystyle\frac{2\alpha}{\rho_0}-
          \displaystyle\frac{2}{\gamma}C_0 ,\\ [0.3cm]
  V_n &=& -\displaystyle\frac{1}{\rho_0}\left(
           \displaystyle\frac{2}{\gamma}\sum\limits^{n}_{l=0}C_l \rho_{n-l}+\sum\limits^{n-1}_{l=0}V_l \rho_{n-l}
  \right),~~~~~~n\ge1 .
  \end{array}
  \right.
  \label{eq_coef_Vn}
 \end{equation}

\section{DISPERSION RELATION OF SAUSAGE WAVES}
\label{sec_DR}
\subsection{Dispersion Relations for Arbitrary Transverse Profiles}

Our derivation of the DR of sausage waves starts with linearizing the ideal MHD equations.
Let $\delta \rho$, $\delta \vec{v}$, $\delta \vec{B}$, and $\delta p$ denote
    the perturbations to the density, velocity, magnetic field and pressure, respectively.
We then proceed by Fourier-decomposing any perturbed value $\delta f(r, z;t)$ as
\begin{eqnarray}
\label{eq_Fourier_ansatz}
  \delta f(r,z;t)={\rm Re}\left\{\tilde{f}(r)\exp\left[-i\left(\omega t-kz\right)\right]\right\}~.
\end{eqnarray}
Now with the definition of the Fourier amplitude for the Lagrangian displacement $\tilde{\xi}_r = i\tilde{v}_r/\omega$,
    one finds that $\tilde{\xi}_r$ is governed by
\begin{equation}
  \displaystyle\frac{{\rm d}}{{\rm d}r}\left[\displaystyle\frac{\rho(c_{\rm s}^2+v^2_{\rm A})(\omega^2-k^2c_{\rm T}^2)}{r(\omega^2-k^2c_{\rm s}^2)}
  \displaystyle\frac{{\rm d}y}{{\rm d}r}\right]+\displaystyle\frac{\rho(\omega^2-k^2v_{\rm A}^2)}{r}y=0 ,
\label{eq_govern_y}
\end{equation}
    where $y \equiv r \tilde{\xi}_r$ turns out to be more convenient to work with
    \citep[see e.g.,Eq.~16 in][]{1992SoPh..138..233G}.

The solution to Eq.~(\ref{eq_govern_y}) in a uniform medium is well-known \citep[e.g.,][]{1983SoPh...88..179E,1986SoPh..103..277C}.
To find its solution in the nonuniform transition layer (TL),
    {we capitalize on the fact that
    sausage modes do not resonantly couple to slow or torsional Alfv\'en waves
    in the examined equilibrium configuration where tubes are straight and aligned with the equilibrium magnetic field~\citep[e.g.,][]{2011SSRv..158..289G}.
This resonant coupling does not occur even if sausage wave frequencies fall in the Alfv\'en or cusp continuum.
Mathematically speaking, this means that the perturbation equation does not involve genuine singularities and its solution can be found with
    an approach based on regular series expansion.
Take the simpler cold MHD case where slow waves disappear.
For trapped sausage waves, the real-valued $\omega$ can indeed match $k v_{\rm A}$ at some location $r_{\rm A}$ in the TL, making the equation governing the
    Eulerian perturbation of total pressure apparently singular there (see Eq.~4 in \citeauthor{2013ApJ...777..158S}~\citeyear{2013ApJ...777..158S}, hereafter S13).
Proceeding with a singular-expansion-based approach, S13 showed that actually neither the total pressure perturbation
    nor the Lagrangian displacement is singular at $r_{\rm A}$.
We showed in Appendix C of \citet{2016SoPh..291..877G} that the approach adopted by S13 yields results identical to what we found
    with a regular-expansion-based method, the latter approach having the advantage that there is no need to
    find the specific location of $r_{\rm A}$ iteratively.
For leaky sausage waves, we noted that this regular-expansion-based method is more appropriate.
In this case, the perturbation equations do not contain any singularity because
    the real part of the longitudinal phase speed ${\rm Re}(\omega/k)$ exceeds $v_{\rm Ae}$,
    which in turn is larger than the Alfv\'en speed in the TL.
Despite this difference, we stress that the approach in S13 was intended to treat wave modes that are evanescent outside straight tubes
    in the general case with arbitrary azimuthal wavenumbers.
And a singular series expansion is necessary for handling wave modes with azimuthal wavenumbers different from zero.}

{In practice, the solution in the TL is found in the following way once a choice for the density and temperature profiles
    is made.
(Note that in view of applications to QPPs in flare loops,
    we choose $v_{\rm Ae} > v_{\rm Ai} > c_{\rm si} > c_{\rm se}$ and
    $c_{\rm s}<v_{\rm A}$ in the TL.)
We first reformulate Eq.~(\ref{eq_govern_y}) such that $\rho$ does not appear, which is necessary for us to streamline our numerical evaluation.
To this end, we note that Eq.~(\ref{eq_forcebalance_rhovacs}) allows $\rho$ to be expressed as
\begin{equation}
  \rho = \frac{2\alpha}{v_{\rm A}^2+2c_{\rm s}^2/\gamma}~. \nonumber
\end{equation}
Now that $\alpha$ is a constant, Eq.~(\ref{eq_govern_y}) is then equivalent to
\begin{equation}
  \displaystyle\frac{{\rm d}}{{\rm d}r}\left[\displaystyle\frac{(c_{\rm s}^2+v^2_{\rm A})(\omega^2-k^2c_{\rm T}^2)}{r(v_{\rm A}^2+2c_{\rm s}^2/\gamma)(\omega^2-k^2c_{\rm s}^2)}
  \displaystyle\frac{{\rm d}y}{{\rm d}r}\right]+\displaystyle\frac{(\omega^2-k^2v_{\rm A}^2)}{r(v_{\rm A}^2+2c_{\rm s}^2/\gamma)}y=0~,
\label{eq_govern_y_TL}
\end{equation}
   which is solved by the following procedure.
First, the coefficients $\rho_n$ and $T_n$ in the expansions of $\rho$ and $T$ are readily evaluated with Eq.~(\ref{eq_rho_coef}).
Second, the coefficients ($C_n$ and $V_n$) that appear in the expansions of $c_{\rm s}^2$ and $v_{\rm A}^2$ are found with Eqs.~(\ref{eq_coef_Cn})
    and (\ref{eq_coef_Vn}), respectively.
Third, given that Eq.~(\ref{eq_govern_y_TL}) is singularity-free, its solution
}
    can then be expressed as linear combinations of two linearly independent solutions, $y_1$ and $y_2$,
    that are regular series expansions about $x \equiv r-R =0$.
In other words,
\begin{eqnarray}
    y_1(x) = \sum_{n=0}^\infty a_n x^n~, \hspace{0.2cm}
    y_2(x) = \sum_{n=0}^\infty b_n x^n~.
\label{eq_def_y1y2_expansion}
\end{eqnarray}
Inserting the expansion (\ref{eq_def_y1y2_expansion}) into
    Eq.~(\ref{eq_govern_y}) and employing the expansions of $c_{\rm s}^2$ and $v_{\rm A}^2$,
    we then derive the recurrence relations for coefficients $a_n$ and $b_n$
    by demanding the coefficient of $x^n~(n=0,~1,~2,~\cdots)$ to be zero in the resulting equation.
Without loss of generality, we choose
\begin{equation}
  a_0 = R^2, a_1 = 0,~\mbox{and}~b_0 = 0, b_1 =R .
\label{eq_def_a01_b01}
\end{equation}
The rest of the coefficients, however, are too lengthy to be included here
    and are given in Appendix \ref{sec_app_coef_general} instead.
{One sees from this appendix that we can evaluate $a_n$ and $b_n$, and consequently $y_1$ and $y_2$,
    without the intervention of $\rho_n$.}

Now one finds that $y = r\tilde{\xi}_r$ can be expressed as
\begin{eqnarray}
   y(r)=\left\{
   \begin{array}{ll}
      -\displaystyle\frac{A_{\rm i}\mu_{\rm i}rJ_1(\mu_{\rm i} r)}{\rho_{\rm i}(\omega^2-k^2 v_{\rm Ai}^2)},          & 0         \le r \le r_{\rm i}, \\ [0.4cm]
      A_1y_1(x)+A_2y_2(x),      & r_{\rm i} \le r \le r_{\rm e}, \\ [0.2cm]
      -\displaystyle\frac{A_{\rm e}\mu_{\rm e}rH^{(1)}_1(\mu_{\rm e} r)}{\rho_{\rm e}(\omega^2-k^2 v_{\rm Ae}^2)},        & r \ge r_{\rm e},
   \end{array} \right.
\label{eq_y_solution_entire}
\end{eqnarray}
    where $A_{\rm i}, A_{\rm e}, A_1$ and $A_2$ are arbitrary constants.
Furthermore, $J_n$ and $H_n^{(1)}$ are the $n$-th-order Bessel and Hankel functions
    of the first kind, respectively (here $n=1$).
As for the quantities $\mu_{\rm i, e}$, they are defined as
\begin{equation}
 \mu_{\rm i, e}^2 =\displaystyle\frac{(\omega^2 - k^2v_{\rm Ai, e}^2)(\omega^2 - k^2c_{\rm si, e}^2)}{(c_{\rm si, e}^2+v_{\rm Ai, e}^2)(\omega^2 - k^2c_{\rm Ti, e}^2)}~.
\label{eq_def_mu}
\end{equation}

To derive the DR also requires the explicit expressions for the Eulerian perturbation
    of total pressure $\tilde{p}_{\rm T}$.
It is related to the Lagrangian displacement via
\begin{eqnarray}
  \tilde{p}_{\rm T} = -
     \displaystyle\frac{\rho(c_{\rm s}^2+v_{\rm A}^2)(\omega^2 - k^2c_{\rm T}^2)}{r(\omega^2 - k^2c_{\rm s}^2)}\left(r\tilde{\xi}_r\right)' ,
\label{eq_Fourie_ptot_xi}
\end{eqnarray}
   where the prime $' = {\rm d}/{\rm d} r$.
With the aid of Eq.~(\ref{eq_y_solution_entire}), one finds that
\begin{equation*}
\tilde{p}_{\rm T}(r)=
  \left\{ \begin{array}{ll}
    A_{\rm i}J_0(\mu_{\rm i}r)~,
        & 0 \le r \leq r_{\rm i}, \\[0.3cm]
    A_{\rm e}H^{(1)}_0(\mu_{\rm e}r)~,    & r \ge r_{\rm e},
   \end{array}
   \right.
\end{equation*}
    in the uniform internal and external media.
On the other hand, in the TL it can be expressed as
\begin{eqnarray}
   \tilde{p}_{\rm T}(x)= -
     \displaystyle\frac{\sum\limits^\infty_{l=0}\rho_l x^l\left(\omega^2\sum\limits^\infty_{n=0}C_n x^n+\omega^2\sum\limits^\infty_{n=0}V_n x^n - k^2\sum\limits^\infty_{n=0}C_n x^n\sum\limits^\infty_{j=0}V_j x^j\right)}
     {(x+R)\left(\omega^2 - k^2\sum\limits^\infty_{n=0}C_nx^n\right)}
        \left[ A_1y_1'(x)
               +A_2y_2'(x)
        \right]~.
\end{eqnarray}
Requiring that $y$ and $\tilde{p}_{\rm T}$ be continuous
     at $r=r_{\rm i}$ and $r=r_{\rm e}$ yields four algebraic equations
     governing $[A_0, A_1, A_{\rm i}, A_{\rm e}]$.
For the solutions to be non-trivial, one finds that
\begin{equation}
 \displaystyle\frac{
  \displaystyle\frac{\rho_{\rm i}J_0(\mu_{\rm i}r_{\rm i})(\omega^2-k^2 v_{\rm Ai}^2)}{\mu_{\rm i}r_{\rm i}J_1(\mu_{\rm i}r_{\rm i})}y_1(x_{\rm i})+\Lambda_{\rm i}y'_1(x_{\rm i})
  }
  {
  \displaystyle\frac{\rho_{\rm i}J_0(\mu_{\rm i}r_{\rm i})(\omega^2-k^2 v_{\rm Ai}^2)}{\mu_{\rm i}r_{\rm i}J_1(\mu_{\rm i}r_{\rm i})}y_2(x_{\rm i})+\Lambda_{\rm i}y'_2(x_{\rm i})
  }
  -
  \displaystyle\frac{
  \displaystyle\frac{\rho_{\rm e}H^{(1)}_0(\mu_{\rm e}r_{\rm e})(\omega^2-k^2 v_{\rm Ae}^2)}{\mu_{\rm e}r_{\rm e}H^{(1)}_1(\mu_{\rm e}r_{\rm e})}y_1(x_{\rm e})+\Lambda_{\rm e}y'_1(x_{\rm e})
  }
  {
  \displaystyle\frac{\rho_{\rm e}H^{(1)}_0(\mu_{\rm e}r_{\rm e})(\omega^2-k^2 v_{\rm Ae}^2)}{\mu_{\rm e}r_{\rm e}H^{(1)}_1(\mu_{\rm e}r_{\rm e})}y_2(x_{\rm e})+\Lambda_{\rm e}y'_2(x_{\rm e})
  }
  =0~,
\label{eq_DR}
\end{equation}
    in which $x_{\rm i,e} = \mp l/2$ and
\begin{equation}
 \Lambda_{\rm i,e}=-
     \displaystyle\frac{\sum\limits^\infty_{l=0}\rho_l x^l_{\rm i,e}\left(\omega^2\sum\limits^\infty_{n=0}C_n x^n_{\rm i,e}+\omega^2\sum\limits^\infty_{n=0}V_n x^n_{\rm i,e} - k^2\sum\limits^\infty_{n=0}C_nx^n_{\rm i,e}\sum\limits^\infty_{j=0}V_jx^j_{\rm i,e}\right)}
     {(x_{\rm i,e}+R)\left(\omega^2 - k^2\sum\limits^\infty_{n=0}C_nx_{\rm i,e}^n\right)}~.
 \label{eq_Lambda}
 \end{equation}

Equation~(\ref{eq_DR}) is the DR valid for arbitrary choices
    of the transverse profiles in the TL.
We note that expressing the external solution in terms of $H_0^{(1)}$ and $H_1^{(1)}$ is necessary to
    provide a unified treatment for both trapped and leaky waves.
In fact, the trapped regime results when
    $\arg\mu_{\rm e} = \pi/2$, from which one finds that
    $H_1^{(1)}(\mu_{\rm e} r) = -(2/\pi) K_1(|\mu_{\rm e}| r)$
    where $K_1$ is the first order modified Bessel function of the second kind.
However, in the leaky regime $\mu_{\rm e}$ is complex valued, resulting in an outward energy flux
    that accounts for the apparent wave damping~\citep[see the discussions in][]{1986SoPh..103..277C,2015A&A...581A.130G}.

\subsection{Dispersion Relation for Top-hat Transverse Profiles}
\label{sec_sub_DRtophat}
In the limit $l/R \rightarrow 0$, one expects the DR~(\ref{eq_DR}) to
    recover the well-known result for top-hat profiles.
To show this, we retain only terms to the 0-th order in $l/R$
    and note that $r_{\rm i}\approx r_{\rm e} \approx R$ and $x_{\rm i}\approx x_{\rm e}$.
Now that $\Lambda_{\rm i}\approx\Lambda_{\rm e}$,
     it follows from Eq.~(\ref{eq_DR}) that
\begin{eqnarray*}
\left(\displaystyle\frac{\rho_{\rm i}J_0(\mu_{\rm i}R)(\omega^2-k^2 v_{\rm Ai}^2)}{\mu_{\rm i}J_1(\mu_{\rm i}R)}-\displaystyle\frac{\rho_{\rm e}H^{(1)}_0(\mu_{\rm e}R)(\omega^2-k^2 v_{\rm Ae}^2)}{\mu_{\rm e}H^{(1)}_1(\mu_{\rm e}R)}\right)(a_1 b_0-a_0 b_1)=0 .
\end{eqnarray*}
Since $a_1 b_0 - a_0 b_1$ is not allowed to be zero
    for $y_1(x)$ and $y_2(x)$ to be independent, this equation suggests that
\begin{eqnarray}
\displaystyle\frac{\rho_{\rm i}J_0(\mu_{\rm i}R)(\omega^2-k^2 v_{\rm Ai}^2)}{\mu_{\rm i}J_1(\mu_{\rm i}R)}=\displaystyle\frac{\rho_{\rm e}H^{(1)}_0(\mu_{\rm e}R)(\omega^2-k^2 v_{\rm Ae}^2)}{\mu_{\rm e}H^{(1)}_1(\mu_{\rm e}R)} ,
\label{eq_DR_tophat}
\end{eqnarray}
   which is the DR for top-hat profiles \citep[e.g.,][]{1986SoPh..103..277C,2007AstL...33..706K}.
From Eq.~(\ref{eq_DR_tophat}) follows that
    the critical wavenumber $k_{\rm c}$ for the {lowest} order sausage mode can be expressed as
\begin{equation}\label{eq_kc_tophat}
  k_{\rm c}R=j_{0,0}\displaystyle\sqrt{\frac{\left(c_{\rm si}^2+v_{\rm Ai}^2\right)\left(v_{\rm Ae}^2-c_{\rm Ti}^2\right)}{\left(v_{\rm Ae}^2-c_{\rm si}^2\right)\left(v_{\rm Ae}^2-v_{\rm Ai}^2\right)}}
\end{equation}
    {with $j_{0,0} = 2.4048$ being} the first zero of $J_0$.

\section{NUMERICAL RESULTS}
\label{sec_para_study}

\subsection{Prescriptions for Transition Layer Profiles and Method of Solution}
\label{sec_sub_solmethod}
\begin{figure}
\centering
 \includegraphics[width=0.9\columnwidth]{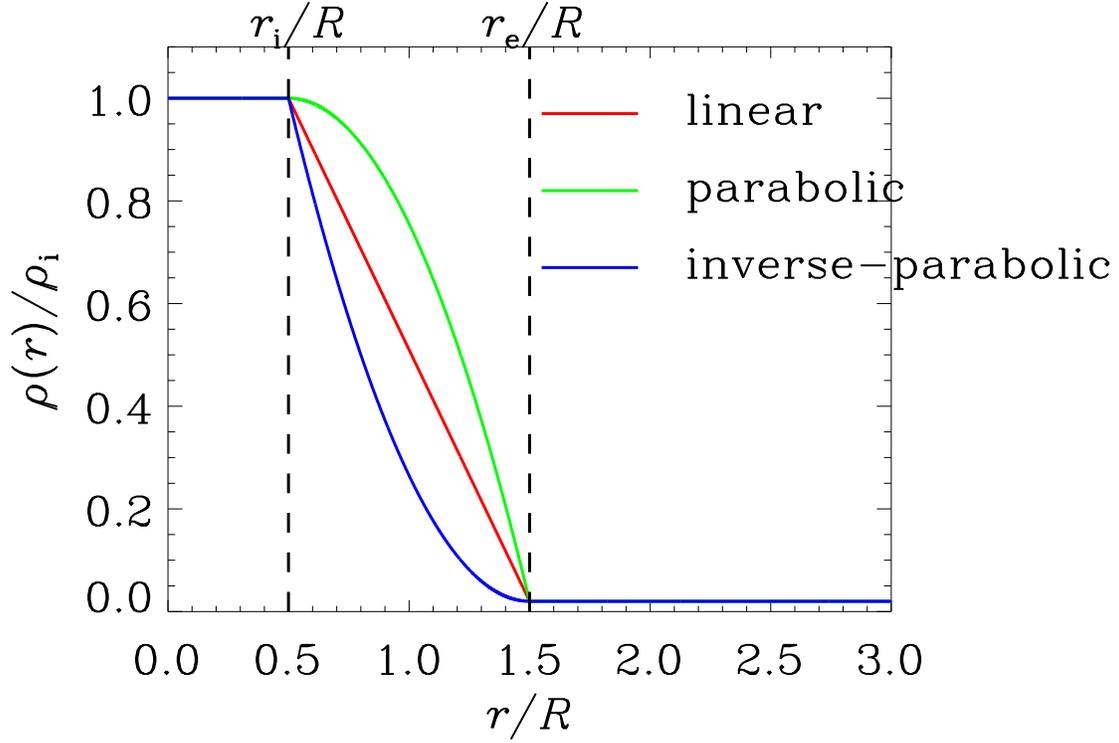}
 \caption{
 Illustration of profile prescriptions using
     transverse equilibrium density profiles as an example.
 These profiles differ only in a transition layer
     sandwiched between the internal (with a uniform density $\rho_{\rm i}$)
     and external (with a uniform density $\rho_{\rm e}$) portions.
 This transition layer is of width $l$, and is located between
     $r_{\rm i}= R-l/2$ and $r_{\rm e}=R+l/2$, with $R$ being the mean tube radius.
 Three different profile prescriptions are adopted as labeled, and are given by
     Eq.~(\ref{eq_TL_profile}).
For illustration purposes, $l$ is chosen to be $R$,
     and $\rho_{\rm i}/\rho_{\rm e}$ is chosen to be $50$.
}
 \label{fig_illus_profile}
\end{figure}

When deriving the DR (Eq.~\ref{eq_DR}), we imposed no restrictions on the profiles for $T(r)$ and $\rho (r)$ in the transition layer
     except that the sound speed is not larger than the Alfv\'en speed therein.
However, in general the transcendental DR is not analytical tractable.
For its numerical evaluation, one has to choose a prescription for $T(r)$ and $\rho(r)$, or
     ${\cal F}(\epsilon_{\rm i}, \epsilon_{\rm e}; r)$ to be precise.
To this end the following choices are adopted,
\begin{eqnarray}
\label{eq_TL_profile}
   {\cal F}(\epsilon_i, \epsilon_e; r) =
   \left\{
   \begin{array}{ll}
   \epsilon_{\rm i}-\displaystyle\frac{\epsilon_{\rm i}-\epsilon_{\rm e}}{l}\left(r-R+\displaystyle\frac{l}{2}\right),   & {\rm linear},
   \\[0.3cm]
   \epsilon_{\rm i}-\displaystyle\frac{\epsilon_{\rm i}-\epsilon_{\rm e}}{l^2}\left(r-R+\displaystyle\frac{l}{2}\right)^2,& {\rm parabolic},\\[0.3cm]
   \epsilon_{\rm e}-\displaystyle\frac{\epsilon_{\rm e}-\epsilon_{\rm i}}{l^2}\left(r-R-\displaystyle\frac{l}{2}\right)^2,    & {\rm inverse-parabolic} .
   \end{array}
   \right.
\end{eqnarray}
We note that these profiles have been extensively used in examinations of kink \citep[e.g.,][and references therein]{2013ApJ...777..158S}
    and sausage (paper I) waves in coronal tubes, albeit in the zero-$\beta$ limit where the temperature profile is irrelevant.
Figure~\ref{fig_illus_profile} uses the transverse density distribution as an example to show
    the different choices for ${\cal F}(\epsilon_{\rm i}, \epsilon_{\rm e}; r)$,
    where we arbitrarily choose $\rho_{\rm i}/\rho_{\rm e} =50$ and $l/R=1$.

Let us focus on fundamental, standing, fast sausage modes of the lowest order, given their importance in accounting for QPPs.
This means that we numerically solve the DR (Eq.~\ref{eq_DR})
    for complex-valued angular frequencies $\omega$ at a given real-valued longitudinal wavenumber $k$,
    which is related to the tube length $L$ via $k = \pi/L$.
To do this requires that the infinite series expansion in Eq.~(\ref{eq_def_y1y2_expansion}) be truncated
    such that only terms up to $n=N$ are kept.
A value of $N=101$ is adopted for all the numerical results to be presented,
    and we made sure that choosing an even larger $N$ does not introduce any discernible difference.
Besides, looking at the coefficients given in Appendix \ref{sec_app_coef_general},
    one sees that 4-fold summations are involved.
This is quite time consuming.
However, for the profiles we choose in Eq.~(\ref{eq_TL_profile}), it is possible to reduce the computational load by
    reformulating these coefficients such that only
    2-fold summations need to be evaluated
    (see Appendix~\ref{sec_app_coef_para} for details).

In short, what comes out of the computations is that once a ${\cal F}(\epsilon_{\rm i}, \epsilon_{\rm e}; r)$ is chosen,
    the dimensionless angular frequency $\omega R/v_{\rm Ai}$ can be formally expressed as
\begin{eqnarray}
    \frac{\omega R}{v_{\rm Ai}} = {\cal G}\left(\frac{L}{R}, \frac{l}{R}, \frac{\rho_{\rm i}}{\rho_{\rm e}}, \beta_{\rm i}, \beta_{\rm e}\right) ,
\label{eq_omega_formal}
\end{eqnarray}
    where $\beta_{\rm i,e} = 2 c_{\rm si,e}^2/(\gamma v_{\rm Ai,e}^2)$ (see Eq.~\ref{eq_def_beta}).
Note that the $L/R$-dependence in Eq.~(\ref{eq_omega_formal}) comes from the dependence on $k$.
The periods $P$ and damping times $\tau$ of sausage modes simply follow from the definitions
    $P=2\pi/{\rm Re}(\omega)$ and $\tau = 1/|{\rm Im}(\omega)|$.
Given that the loops we examine are embedded in a background corona,
    we fix $\beta_{\rm e}$ at $0.01$.
Experimenting with an arbitrarily chosen subset of the numerical results,
    we found that using a smaller $\beta_{\rm e}$ brings forth changes of less than $1\%$.

\subsection{Effects of a finite beta}
\label{sec_sub_beta_effects}
\begin{figure}
\centering
\includegraphics[width=0.5\columnwidth]{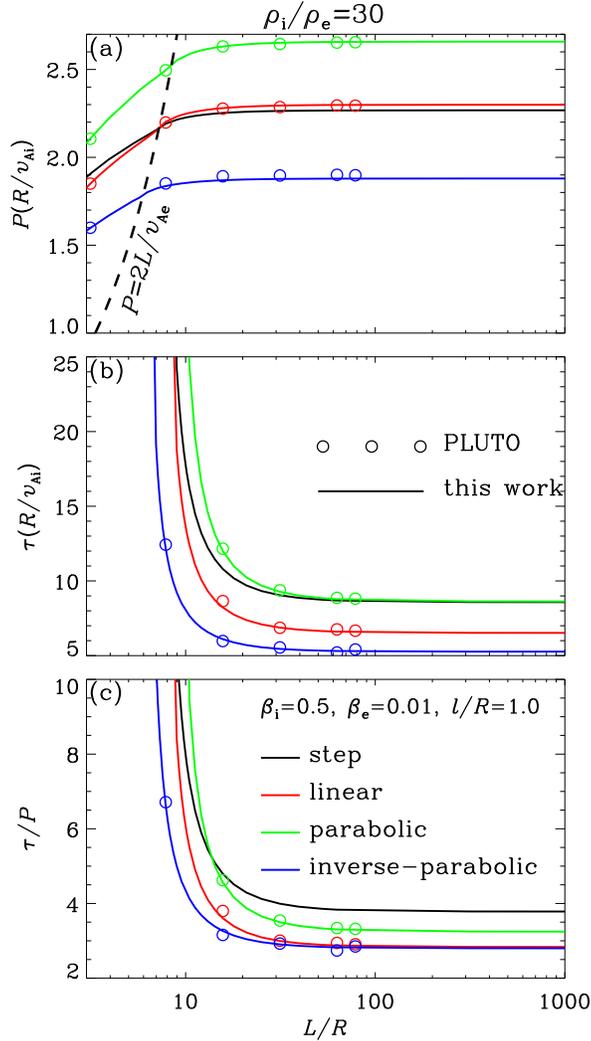}
 \caption{
 Dispersive properties of fast sausage modes in transversally continuous tubes with a finite plasma $\beta$.
 The periods ($P$, panel a), damping times ($\tau$, panel b)
    and damping-time-to-period ratios ($\tau/P$, panel c)
    are plotted as functions of the length-to-radius ratio $L/R$.
 The curves in various colors represent the results for a number of transverse profile prescriptions as labeled,
    for which the density contrast $\rho_{\rm i}/\rho_{\rm e}=30$,
    the transverse lengthscale $l=R$,
    and the internal (external) plasma $\beta$ is $0.5$ ($0.01$).
 For comparison, the black curves represent the results for top-hat profiles
    (or equivalently $l/R \rightarrow 0$).
 The black dashed curve in (a) represents $P=2L/v_{\rm Ae}$ and separates
     the trapped (to its left) from leaky (right) regime.
 {In addition, the open circles represent the periods and damping times found by solving the time-dependent ideal MHD equations,
     an approach independent from the eigenmode analysis.}
 }
 \label{fig_Ptau_vs_L}
\end{figure}

To start, let us examine a solution to the DR as given in Fig.~\ref{fig_Ptau_vs_L},
     where $P$ and $\tau$ in units of the transverse Alfv\'en time $R/v_{\rm Ai}$
     are shown as a function of the length-to-radius ratio $L/R$
     for a combination $[{l}/{R}, {\rho_{\rm i}}/{\rho_{\rm e}}, \beta_{\rm i}]$
     of $[1, 30, 0.5]$.
The results for different profiles are given in different colors as labeled in Fig.~\ref{fig_Ptau_vs_L}c.
For comparison, the black solid curves represent the results for the corresponding top-hat profile ($l/R =0$).
In Fig.~\ref{fig_Ptau_vs_L}a, the black dash-dotted line represents $P = 2L/v_{\rm Ae}$, which separates
     the trapped (to its left) from the leaky (right) regime.
{In addition, the open circles represent the periods and damping times obtained with an initial-value-problem (IVP) approach
     by solving the full ideal MHD equations with the PLUTO code,
     which is detailed in Appendix \ref{sec_app_ivp} and independent from the eigenmode analysis.
One sees that the open circles agree remarkably well with the solid curves, thereby validating our eigenmode analysis.
Furthermore, Fig.~\ref{fig_Ptau_vs_L}a indicates that}
     $P$ tends to increase with $L/R$ in the trapped regime
     and rather rapidly settles to a constant in the leaky regime.
On the other hand, being infinite in the trapped regime, $\tau$ decreases with $L/R$
     and tends to some constant when $L/R$ is sufficiently large {(see Fig.~\ref{fig_Ptau_vs_L}b)}.
These features, regardless of the profile choices,
     are reminiscent of the zero-$\beta$ results
     as given by Fig.~3 in \citet{2012ApJ...761..134N}
     and Fig.~2 in \citet{2015SoPh..290.2231C},
     although different density profiles were chosen therein.
This means that a finite $\beta$ does not affect the qualitative behavior of sausage modes,
     as far as the overall dependence of their periods and damping times on tube length is concerned.
In particular, a critical $L/R$ (or equivalently a critical longitudinal wavenumber $k_{\rm c}$) still exists.

\begin{figure}
\centering
\includegraphics[width=.95\columnwidth]{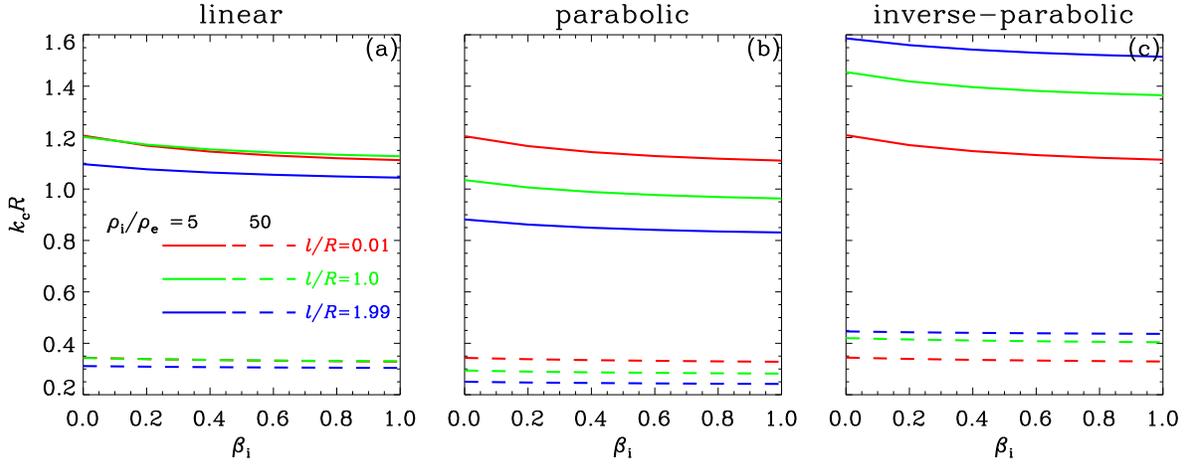}
 \caption{
  Dependence on the internal plasma beta ($\beta_{\rm i}$) of critical wavenumbers $k_{\rm c}$
      of fast sausage modes in transversally continuous tubes.
  Three different transverse profile prescriptions are given in different panels.
  A number of combinations for the density contrast $\rho_{\rm i}/\rho_{\rm e}$
      and transverse lengthscale $l/R$
      are examined as labeled.
  The external plasma beta is fixed at $0.01$.
  {Note that in panel (a), the red dashed curve
       can be hardly seen because it almost coincides with the green one.}
  }
 \label{fig_kc_vs_betai}
\end{figure}

One way to bring out the quantitative influence of a finite beta is to examine how
     $k_{\rm c}$ changes with $\beta_{\rm i}$.
This is shown in Fig.~\ref{fig_kc_vs_betai} where
     the results for different profiles are given in different panels.
A number of combinations $[{l}/{R}, {\rho_{\rm i}}/{\rho_{\rm e}}]$ are adopted
     and given by different colors and linestyles.
Consider Fig.~\ref{fig_kc_vs_betai}a first, which pertains to linear profiles.
One sees that by far the most important factor that influences $k_{\rm c}$ is the density contrast:
     $k_{\rm c}$ decreases substantially with $\rho_{\rm i}/\rho_{\rm e}$.
This is understandable as one intuitively expects that coronal tubes
     become more efficient in trapping sausage modes
     when the density contrast increases.
In addition, $k_{\rm c}$ tends to decrease with the transverse lengthscale $l/R$, even though
     $k_{\rm c}$ hardly varies when $l/R \lesssim 1$ for linear profiles.
Somehow $k_{\rm c}$ is not sensitive to $\beta_{\rm i}$, which is particularly true for large density contrasts.
Even for a $\rho_{\rm i}/\rho_{\rm e}$ as small as $5$ (the solid curves),
     $k_{\rm c}$ at $\beta_{\rm i} =1$ is smaller than its value attained in the zero-$\beta$ case
     by no more than $7.9\%$ for the transverse lengthscales considered.
One may understand this behavior by examining the case where $l/R\rightarrow 0$ (the red curves),
     for which $k_{\rm c}$ agrees exactly with the analytical expectation given by Eq.~(\ref{eq_kc_tophat}).
Reformulating Eq.~(\ref{eq_kc_tophat}) in terms of dimensionless values and focusing on the lowest-order mode, one finds that
\begin{eqnarray}
     k_{\rm c} R = 2.4048 \sqrt{-1+\frac{\rho_{\rm ie}^2 (1+\beta_{\rm i})^2}
         {[\rho_{\rm ie}(1+\beta_{\rm i})-(1+\beta_{\rm e})][\rho_{\rm ie}(1+\beta_{\rm i})-(\gamma\beta_{\rm i}/2)(1+\beta_{\rm e})]}} ,
     \label{eq_kc_tophat_dmless}
\end{eqnarray}
     where $\rho_{\rm ie} = \rho_{\rm i}/\rho_{\rm e}$.
Given that $\beta_{\rm e} \ll 1$, one finds that $k_{\rm c}$ can be approximated to within $\sim 10\%$ by
\begin{eqnarray}
     k_{\rm c} R \approx 2.4048 \sqrt{\frac{1+\gamma\beta_{\rm i}/2}{\rho_{\rm ie}(1+\beta_{\rm i})}} ,
     \label{eq_kc_apprx_tophat}
\end{eqnarray}
     when $\rho_{\rm ie}\gtrsim 5$.
Equation~(\ref{eq_kc_apprx_tophat}) suggests that the dependence on $\beta_{\rm i}$ is largely offset
     by the appearance of the square root and
     the fact that $\gamma/2$ is close to $1$.
Actually, the weak $\beta_{\rm i}$-dependence of $k_{\rm c}$ for top-hat profiles was already shown
     by \citet{2009A&A...503..569I}, even though a slab geometry was adopted there.
It is no surprise to see the same weak dependence for a cylindrical geometry given that
      for top-hat profiles, $k_{\rm c}$ for the two geometries differ by only a numerical factor.
What Fig.~\ref{fig_kc_vs_betai}a suggests is that for continuous profiles in a linear form,
     this weak dependence persists.
Now move on to Figs.~\ref{fig_kc_vs_betai}b and \ref{fig_kc_vs_betai}c, which pertain to
     the parabolic and inverse-parabolic profiles, respectively.
One sees that $k_{\rm c}$ also only weakly depends on $\beta_{\rm i}$.
Furthermore, once again $k_{\rm c}$ is most sensitive to the density contrast $\rho_{\rm i}/\rho_{\rm e}$.
However, one thing peculiar is that for the inverse-parabolic profile
     $k_{\rm c}$ tends to increase with $l/R$, which is opposite to
     the tendency for linear and parabolic profiles.
Intuitively speaking, one would expect that coronal tubes will become
     less efficient in wave trapping when their boundaries become more diffusive,
     and hence a larger $k_{\rm c}$.
The reason why this expectation does not take place for linear and parabolic profiles
     may be attributed to the transverse mass distribution.
If evaluating $M = \int_0^{2R} \rho_{r} r dr$, the mass per unit longitudinal length,
     one finds that $M$ tends to decrease with $l/R$  for inverse-parabolic profiles, meaning that
     the tube becomes effectively thinner and $R$ may overestimate the effective radius $R_{\rm eff}$.
As a result, with increasing $l/R$, the curves in Fig.~\ref{fig_kc_vs_betai}c may be lowered if $k_{c} R_{\rm eff}$
     is plotted instead of $k_{c} R$.
In contrast, for linear and parabolic profiles, $M$ tends to increase with $l/R$,
     meaning that $R$ may underestimate $R_{\rm eff}$
     and the curves in Figs.~\ref{fig_kc_vs_betai}a and \ref{fig_kc_vs_betai}b
     may be shifted upwards if $k_{c} R_{\rm eff}$ is plotted.
Replacing $R$ with a proper $R_{\rm eff}$ may bring the results
     for different $l/R$ closer to the intuitive expectation,
     however this is beyond the scope of the present manuscript.

\begin{figure}
\centering
\includegraphics[width=.95\columnwidth]{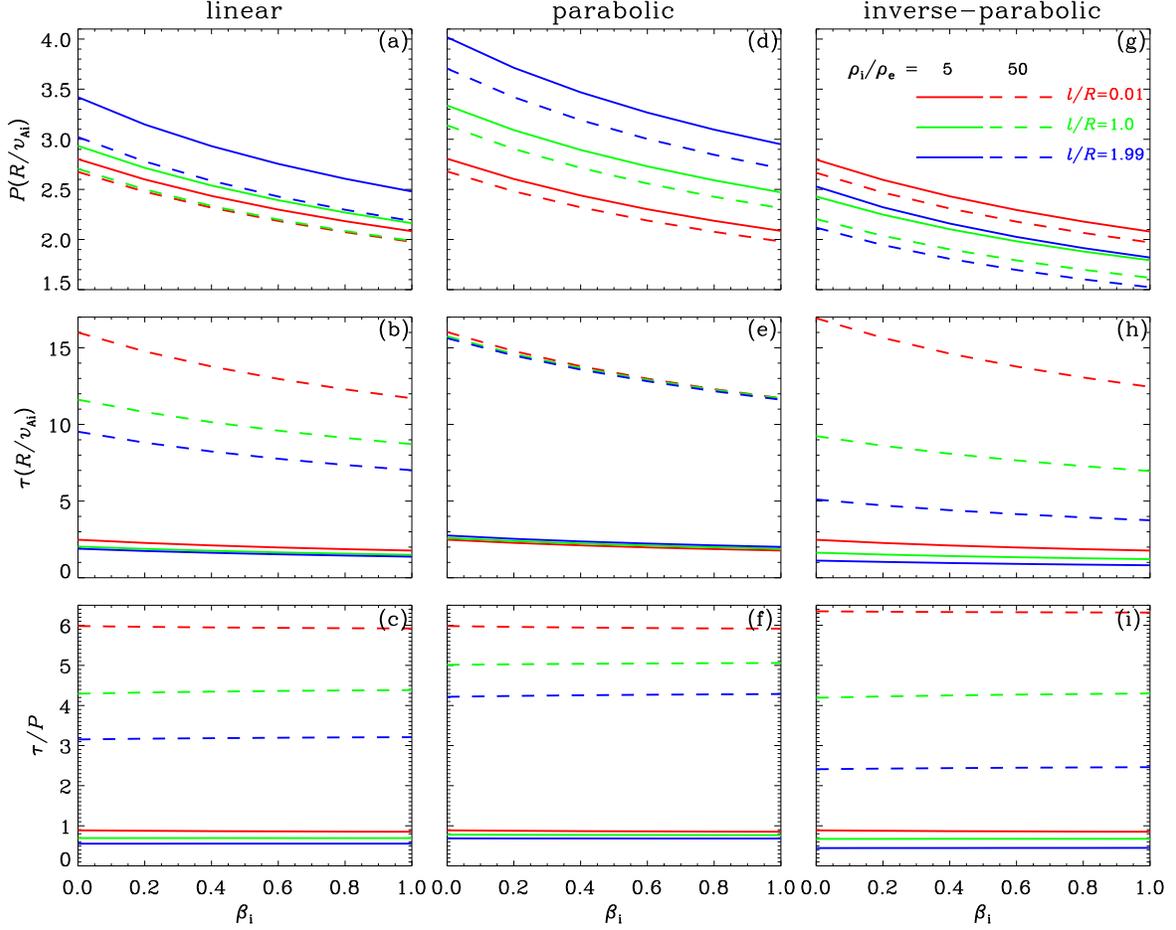}
 \caption{
  Dependence on the internal plasma beta ($\beta_{\rm i}$)
     of the (a) periods $P$, (b) damping times $\tau$,
     and (c) damping-time-to-period ratios $\tau/P$
     of fast sausage modes in transversally continuous tubes with a length-to-radius ratio $L/R$ of $1000$.
  Here $P$ and $\tau$ are in units of the transverse Alfv\'en time $R/v_{\rm Ai}$ (see Eq.~\ref{eq_omega_formal}).
  Three different transverse profile prescriptions are given in different columns.
  A number of combinations for the density contrast $\rho_{\rm i}/\rho_{\rm e}$
      and transverse lengthscale $l/R$ are examined as labeled.
  The external plasma beta is fixed at $0.01$.
  }
 \label{fig_saturated_betai_vai}
\end{figure}

The $\beta_{\rm i}$ dependence can be also brought out by
     examining how the periods $P$ and damping times $\tau$ vary.
A simple way to do this is to examine the limit where $L/R \rightarrow \infty$ (or equivalently $k\rightarrow 0$),
     given that neither $P$ nor $\tau$ depends on $L/R$ for sufficiently large $L/R$.
Figure~\ref{fig_saturated_betai_vai} presents the values of $P$ and $\tau$ thus derived
     for different profiles and for a number of choices for $\rho_{\rm i}/\rho_{\rm e}$ and $l/R$ as labeled.
Note that the ratio $\tau/P$ is also plotted in the bottom row
     since it is a better measure of the signal quality.
Examining this row, one sees that regardless of profile prescriptions,
     $\tau/P$ tends to increase when $\rho_{\rm i}/\rho_{\rm e}$ increases
     or $l/R$ decreases.
This is expected since oscillating coronal tubes will be less efficient in emitting fast waves
     when they become more distinct from their surrounding fluids.
Actually, this also makes defining a proper $R_{\rm eff}$ less urgent because in place of $k_{\rm c}$,
     one may adopt $(\tau/P) (k\rightarrow 0)$ to measure the capability
     for coronal tubes to trap sausage wave energy.
Now consider the first two rows.
One finds that $P$ and $\tau$ tend to decrease with $\beta_{\rm i}$, the tendency being
     particularly pronounced for relatively small values of $\rho_{\rm i}/\rho_{\rm e}$.
Take the case where $\rho_{\rm i}/\rho_{\rm e}=5$ and $l/R=1$ for instance.
From the solid green curves one sees that $P v_{\rm Ai}/R$ at $\beta_{\rm i} = 1$ reads $2.16$ ($2.47$, $1.79$)
     for the linear (parabolic, inverse-parabolic) profile,
      while it attains $2.93$ ($3.34$, $2.43$) when $\beta_{\rm i} = 0$.
In relative terms, this is $26.3\%$ ($26.1\%$ , $26.3\%$) smaller.
However, this rather sensitive $\beta_{\rm i}$-dependence is not seen in $\tau/P$,
     as indicated by the bottom row.
One naturally wonders whether the rather strong dependence of $P$ and $\tau$ comes simply from
     the fact that they are measured in units of the transverse Alfv\'en time $R/v_{\rm Ai}$.

\begin{figure}
\centering
\includegraphics[width=.9\columnwidth]{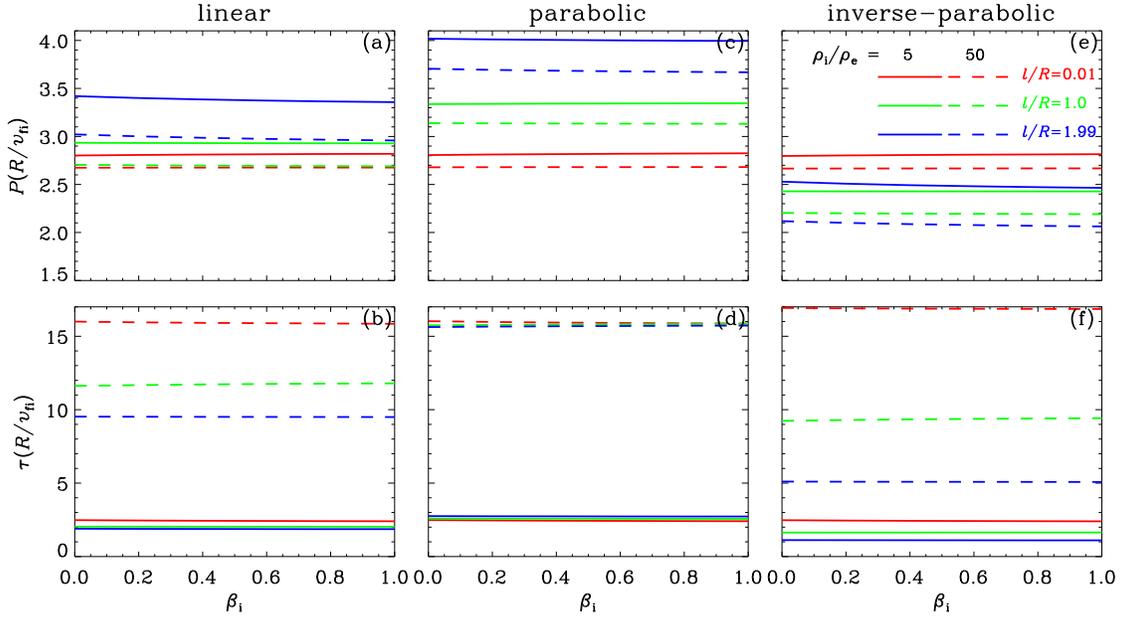}
 \caption{
 Similar to the first two rows in Fig.~\ref{fig_saturated_betai_vai}
     except that $P$ and $\tau$ are in units of the transverse fast time $R/v_{\rm fi}$
     (see Eq.~\ref{eq_omega_formal_in_vfi}).
 }
 \label{fig_saturated_betai_vfi}
\end{figure}

Figure~\ref{fig_saturated_betai_vfi} is essentially the same as the first two rows of Figure~\ref{fig_saturated_betai_vai},
     the only difference is that now $P$ and $\tau$ are expressed in units of the transverse fast time $R/v_{\rm fi}$.
Now one sees that the strong $\beta_{\rm i}$ dependence disappears.
Still take the case where $\rho_{\rm i}/\rho_{\rm e}=5$ and $l/R=1$ for instance.
From the solid green curves one sees that $P v_{\rm fi}/R$ at $\beta_{\rm i} = 1$ reads $2.93$ ($3.35$, $2.43$)
     for the linear (parabolic, inverse-parabolic) profile,
     which is almost identical to the values
     attained when $\beta_{\rm i} = 0$.
Why does this happen?
{Let $\omega_0$ denote the angular frequency attained when $kR \propto R/L \rightarrow 0$.
To understand the insensitivity to $\beta_{\rm i}$ of $P v_{\rm fi}/R$ and $\tau v_{\rm fi}/R$,
     it will then be informative to derive a compact expression for $\omega_0$.
However, it is not straightforward to do so in the general sense given the complexity of the DR (Eq.~\ref{eq_DR}).
Nonetheless, it is easy to show that for top-hat profiles, $\omega_0$ satisfies the following relation
\begin{equation}
   \displaystyle\frac{\rho_{\rm i}}{\rho_{\rm e}}
   =\displaystyle\frac{J_1\left(\omega_0R/v_{\rm fi}\right)}
   {J_0\left(\omega_0R/v_{\rm fi}\right)}
   \displaystyle\frac{H^{(1)}_0\left(\omega_0R/v_{\rm fe}\right)}
   {H^{(1)}_1\left(\omega_0R/v_{\rm fe}\right)}\displaystyle\frac{v_{\rm fe}}{v_{\rm fi}}~,
\label{eq_k0_tophat}
\end{equation}
   which can be found by simply letting $k = 0$ in Eq.~(\ref{eq_DR_tophat}).
Here $v_{\rm fe}$ denotes the external fast speed.
In this simpler case, \citet{1975IGAFS..37....3Z} derived an approximate solution to Eq.~(\ref{eq_k0_tophat})
    for fast sausage modes when the density contrast $\rho_{\rm i}/\rho_{\rm e} \gg 1$.
When expressed in terms of $P$ and $\tau$, this solution reads \citep[][Eqs.~6 and 7]{2007AstL...33..706K}
\begin{equation}
P_{k=0}   \approx \frac{2\pi}{j_{0, 0}} \frac{R}{v_{\rm fi}} = \frac{2.62 R}{v_{\rm fi}}, \hspace{0.5cm}
\tau_{k=0} \approx \frac{P_{k=0}}{\pi^2} \frac{\rho_{\rm i}}{\rho_{\rm e}}~,
\label{eq_Ptau_k0_tophat}
\end{equation}
   which suggests that neither $P$ nor $\tau$ depends on plasma $\beta$ for top-hat profiles
   in the thin-tube limit ($k\rightarrow 0$ or equivalently $L/R \rightarrow \infty$).
Figure~\ref{fig_saturated_betai_vfi} indicates that this insensitivity to plasma $\beta$ in the thin-tube limit persists even when
   the equilibrium parameters are transversally distributed in a continuous manner.
}

{Apart from mathematical reasons, what makes $R/v_{\rm fi}$ special relative to $R/v_{\rm Ai}$?}
This is related to the spatial distributions of the eigen-functions $\tilde{\xi}_r(r)$ and $\tilde{p}_{\rm T}(r)$,
     even though they are not plotted.
These eigen-functions turn out to possess a spatial scale of the order $10 R$, which is substantially smaller
     than the longitudinal lengthscale, the tube length $L$ (taken to be $1000 R$ here).
This means that the effective wavevector for sausage modes is essentially perpendicular to the equilibrium magnetic field,
     hence making $v_{\rm fi}$ more proper than $v_{\rm Ai}$ for describing fast sausage waves.

\begin{figure}
\centering
\includegraphics[width=0.9\columnwidth]{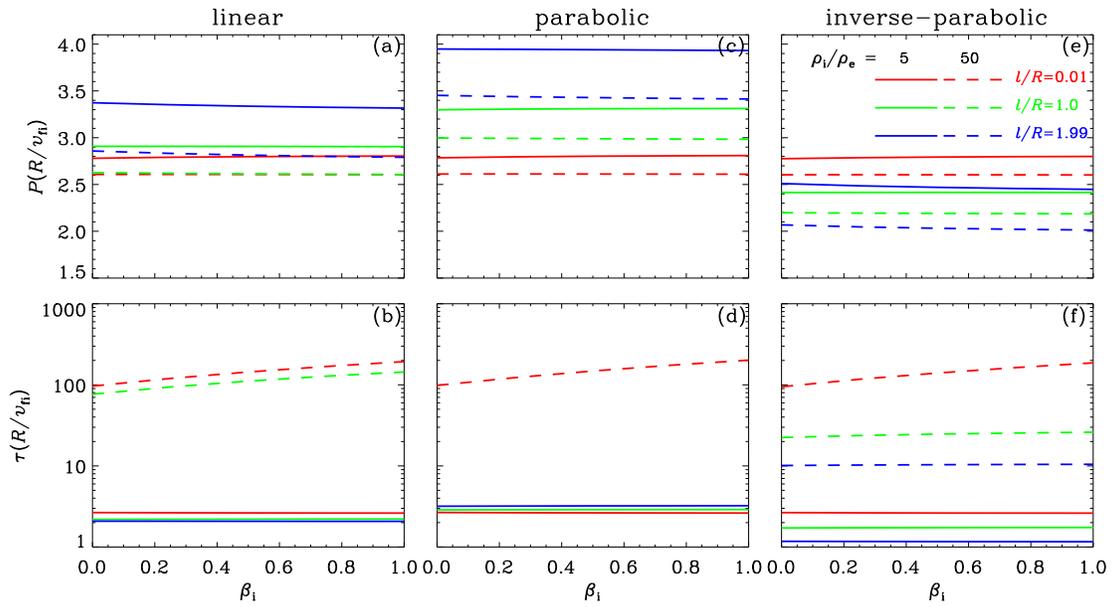}
 \caption{
  Similar to Fig.~\ref{fig_saturated_betai_vfi} but for tubes with a length-to-radius ratio $L/R$ of $10$.
  }
\label{fig_betai_vfi_LR10}
\end{figure}

What Fig.~\ref{fig_saturated_betai_vfi} means is
     that when Eq.~(\ref{eq_omega_formal}) is reformulated as
\begin{eqnarray}
    \frac{\omega R}{v_{\rm fi}} = {\cal H}\left(\frac{L}{R}, \frac{l}{R}, \frac{\rho_{\rm i}}{\rho_{\rm e}}, \beta_{\rm i}, \beta_{\rm e}\right) ,
\label{eq_omega_formal_in_vfi}
\end{eqnarray}
     the function ${\cal H}$ depends only weakly on $\beta_{\rm i}$ for sufficiently thin tubes.
However, does this weak $\beta_{\rm i}$ dependence of ${\cal H}$
     persist for tubes that are not so thin?
Before examining this, we note that for NoRH flare loops hosting sausage modes,
     \citet{2003A&A...412L...7N} found that $L/R \approx 25~\mbox{Mm}/3~\mbox{Mm} = 8.3$,
     while \citet{2015A&A...574A..53K} estimated that
      $L/R \approx 40~\mbox{Mm}/4~\mbox{Mm} = 10$.
On the other hand, for the IRIS flare loop where a global sausage mode was identified,
     \citet{2016ApJ...823L..16T} found that $L/R \approx 30~\mbox{Mm}/3.6~\mbox{Mm} = 8.3$.
Let us take a value of $L/R = 10$ as being representative
     and examine how $P$ and $\tau$ vary at this $L/R$ when $\beta_{\rm i}$ changes.
This is shown in Fig.~\ref{fig_betai_vfi_LR10}, which follows the same format as Fig.~\ref{fig_saturated_betai_vfi}.
Note that in Figs.~\ref{fig_betai_vfi_LR10}b and \ref{fig_betai_vfi_LR10}d,
     not all dashed curves are present because for a density contrast
     as strong as $50$, fast sausage modes are trapped when $l/R=1.99$ ($l/R = 1$ and $1.99$)
     for a linear (parabolic) prescription.
One sees that once again $P$ in units of $R/v_{\rm fi}$ (the upper row)
     depends only weakly on $\beta_{\rm i}$.
That $P$ is primarily determined by the transverse fast time
     is also understandable by comparing the transverse and longitudinal
     lengthscales of the eigenfunctions.
In this case, while $L/R$ is considerably smaller than examined in Fig.~\ref{fig_saturated_betai_vfi},
     the transverse lengthscales are also much smaller (of the order $R$).
The end result is that the wavevector remains largely perpendicular to the equilibrium magnetic field.
Now consider the lower row, where $\tau$ is shown.
One sees that $\tau$ for $\rho_{\rm i}/\rho_{\rm e} = 50$ shows some significant variation for some values of $l/R$
     (see e.g., the green dashed curve corresponding to $l/R = 1$ in Fig.\ref{fig_betai_vfi_LR10}b).
This happens because $L/R$ is not too far from the critical value $(L/R)_{\rm c}$ as determined by $\pi/(k_{\rm c}R)$.
Take the linear profile with $l/R = 1$ for instance.
From the green dashed curve  pertaining to $\rho_{\rm i}/\rho_{\rm e} = 50$ in Fig.~\ref{fig_kc_vs_betai}a,
     one finds that a value of $k_{\rm c}R \approx 0.34$ can be quoted for all $\beta_{\rm i}$.
This results in $(L/R)_{\rm c} \approx 9.24$, which is only marginally smaller than $10$.
On the other hand, if $L/R$ exceeds $(L/R)_{\rm c}$ by say, $50\%$,
     such that the modes are deeper in the leaky regime,
     then the damping time $\tau$ is not sensitive to $\beta_{\rm i}$ any more
     (see e.g., all the solid curves in the lower row).
Actually this can be seen as a rule of thumb from a series of experiments
     that we conducted for $\rho_{\rm i}/\rho_{\rm e}$ between
     {$2$ and $200$} and $L/R$ between $5$ and $100$.
Furthermore, these further computations suggest that $P$ in units of $R/v_{\rm fi}$
     is not sensitive to $\beta_{\rm i}$ for all the chosen profiles,
     be the modes in the trapped or leaky regime.

\begin{figure}
\centering
\includegraphics[width=0.75\columnwidth]{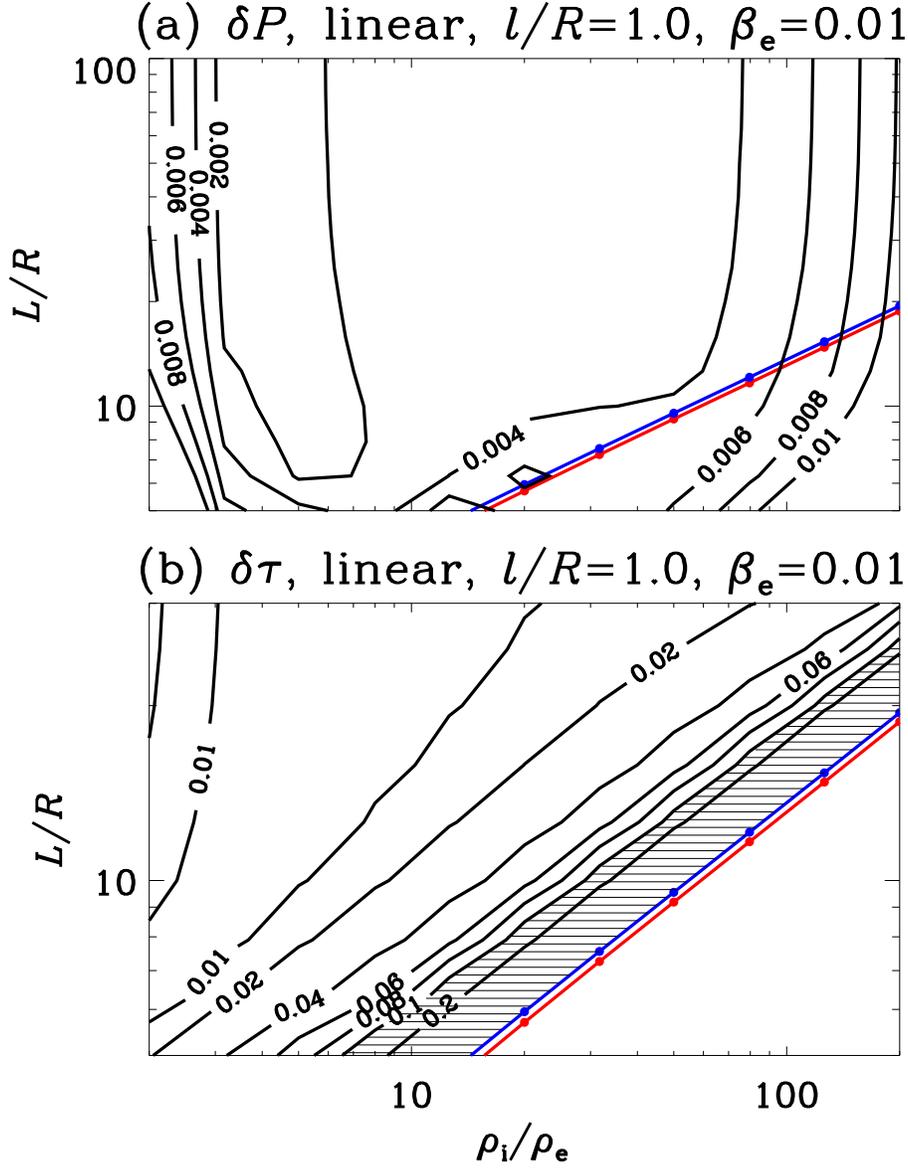}
 \caption{
  Distributions of $\delta P$ and $\delta \tau$ in the space spanned by $\rho_{\rm i}/\rho_{\rm e}$ and $L/R$ for linear profiles with
      $l/R=1$ and $\beta_{\rm e}=0.01$.
  Here $\delta P$ evaluates the maximal difference of the period relative to the cold MHD result
      at a given pair $[\rho_{\rm i}/\rho_{\rm e}, L/R]$ when $\beta_{\rm i}$ varies between $0$ and $1$.
  And $\delta \tau$ is defined in the same fashion for the damping times.
  In addition, the red and blue lines represent the lower and upper limits of the cutoff length-to-radius ratio
      at a given density contrast.
  The hatched area in panel (b) represents where $\delta \tau$ exceeds $10\%$.
  See text for details.
  }
\label{fig_contour_diff_cold}
\end{figure}

{An example of these further computations is given in Figure~\ref{fig_contour_diff_cold} where we adopt a linear profile and fix $[l/R, \beta_{\rm e}]$ at $[1, 0.01]$.
It then follows that the periods $P$ and $\tau$ in units of $R/v_{\rm fi}$ at a given pair of $[\rho_{\rm i}/\rho_{\rm e}, L/R]$
     are functions of $\beta_{\rm i}$ only.
Let them be denoted by $P^{\beta \ne 0}$ and $\tau^{\beta\ne 0}$, respectively.
At a given $[\rho_{\rm i}/\rho_{\rm e}, L/R]$, we then evaluate $P$ and $\tau$ in units of $R/v_{\rm fi}$ in the cold MHD limit by solving
     the corresponding DR (Eq.~17 in paper I).
Note that $v_{\rm Ai}$ equals $v_{\rm fi}$ in this cold MHD case.
Let these values be denoted by $P^{\rm cold}$ and $\tau^{\rm cold}$, respectively.
We now define $\delta P$ and $\delta\tau$ to be the maximal relative difference between the finite-$\beta$ and cold MHD results when $\beta_{\rm i}$
     varies between $0$ and $1$.
In other words,
\begin{eqnarray*}
&& \delta P = {\rm max}\left|\frac{P^{\beta \ne 0}(\beta_{\rm i} \in [0, 1])}{P^{\rm cold}}-1\right| , \\
&& \delta \tau = {\rm max}\left|\frac{\tau^{\beta \ne 0}(\beta_{\rm i} \in [0, 1])}{\tau^{\rm cold}}-1\right|~.
\end{eqnarray*}
Shown in Figure~\ref{fig_contour_diff_cold} are the distributions in the $[\rho_{\rm i}/\rho_{\rm e}, L/R]$ space of (a) $\delta P$ and (b) $\delta \tau$.
In addition, the red and blue lines represent the lower and upper limits of the cutoff $(L/R)_{\rm c}$ when $\beta_{\rm i}$ varies from $0$ to $1$.
Trapped (leaky) modes lie to the right (left) of these lines.
One sees that these two lines are very close to each other, meaning that $(L/R)_{\rm c}$ varies little when $\beta_{\rm i}$ varies.
Furthermore, Fig.~\ref{fig_contour_diff_cold}a indicates that $\delta P$ is consistently less than $1.2\%$ throughout this extensive range of $[\rho_{\rm i}/\rho_{\rm e}, L/R]$.
As to $\delta \tau$, the portion to the right of the red line in Fig.~\ref{fig_contour_diff_cold}b is irrelevant because $\tau$ is identically infinite therein.
One sees that $\delta\tau \gtrsim 10\%$ only in the hatched portion in the immediate vicinity of the blue or red line.
Actually the contour outlining $\delta\tau=10\%$ is almost parallel to the blue or red line, and is consistent with $L/R = 1.5 (L/R)_{\rm c}$.
}

The insensitive dependence on $\beta_{\rm i}$ of $P v_{\rm fi}/R$ is good news from the
     seismological perspective.
When inverting the periods of the trapped sausage modes as measured by
     RoRH~\citep{2003A&A...412L...7N} and IRIS~\citep{2016ApJ...823L..16T},
     one can safely use the much simpler zero-$\beta$ theory presented in paper I,
     and simply see the deduced Alfv\'en speed $v_{\rm Ai}$ as
     the fast speed $v_{\rm fi}$.
However, when leaky modes are measured, caution needs to be exercised:
     given a measured $L/R$ from imaging instruments, in general one cannot safely assume that
     this $L/R$ is far from the critical value $(L/R)_{\rm c}$.
This is largely because one does not know beforehand which profile choice best describes the
     transverse distributions of plasma density and temperature.
And from Figs.~\ref{fig_kc_vs_betai}, one sees that regardless of profile prescriptions,
     $k_{c}R$ and hence $(L/R)_{\rm c}$ can be quite different when $l/R$ varies.
While $\tau$ in units of $R/v_{\rm fi}$ does not change much with $\beta_{\rm i}$ when $L/R$ exceeds
     $(L/R)_{\rm c}$ by $\sim 50\%$, this cannot be guaranteed without knowing $(L/R)_{\rm c}$ beforehand.
Actually this strengthens our suggestion in paper I that in addition to the density contrast and profile steepness,
     the detailed form for describing the transverse density distribution
     also plays an important role in determining the dispersive properties
     of fast sausage modes.

\section{SUMMARY AND CONCLUDING REMARKS}
\label{sec_conclusion}
Standing sausage modes in flare loops have been often invoked to account for quasi-periodic pulsations with
    periods of the order of seconds in the lightcurves of solar flares.
Their measurements, the periods $P$ and $\tau$ in particular, can be used to infer such key information
    as the Alfv\'en speed in key regions where flare energy is released.
Indispensable in this context is a detailed theoretical understanding
    of sausage waves collectively supported by magnetic tubes, for which purpose one usually chooses to
    work in the framework of cold (zero-$\beta$) MHD and/or assumes that the magnetic and plasma parameters
    are transversally structured in a top-hat fashion.
The aim of the present study has been to derive the dispersion relation (DR) for sausage waves that incorporates
    the effects of a continuous transverse structuring and a finite plasma $\beta$,
    the latter being particularly necessary given that $\beta$ in flare loops may reach order unity.
To this end, we adopted linearized ideal MHD equations and modeled coronal loops as straight tubes
    with transverse density and temperature profiles characterized by a transition layer
    sandwiched between a uniform cord and a uniform external medium.
An analytical DR (Eq.~\ref{eq_DR}) was worked out
    by solving the perturbation equations in terms of regular series expansions in the transition layer.
For this to work, we required that sausage waves do not resonantly couple to torsional Alfv\'en waves
    or slow waves.
This is not a severe limitation, and is readily applicable to fast sausage waves in flare loops embedded
    in a background corona.
In addition, this DR is valid for essentially arbitrary distributions of densities and temperatures
    in the transition layer.

In general, we found that $P$ and $\tau$ of standing fast sausage modes depend on a combination of parameters
    $[R/v_{\rm Ai}, L/R, l/R, \rho_{\rm i}/\rho_{\rm e}, \beta_{\rm i}, \beta_{\rm e}]$
    as formally expressed by Eq.~(\ref{eq_omega_formal}).
Here $L$ and $R$ denote the tube length and radius, respectively.
Furthermore, $l$ is the width of the transition layer, $v_{\rm Ai}$ is the
    Alfv\'en speed in the cord, $\rho_{\rm i}/\rho_{\rm e}$ is the density contrast between the loop and its surroundings,
    and $\beta_{\rm i}$ ($\beta_{\rm e}$) represents the plasma $\beta$ in the cord (external medium).
We showed that for the transverse profiles examined, neither $P$ nor $\tau$
    depends on $L/R$ provided that $L/R$ is sufficiently large.
In addition, for a coronal background, the dependence on $\beta_{\rm e}$ disappears as well when
    $\beta_{\rm e}$ is sufficiently small.

The effect of a finite $\beta_{\rm i}$ was quantified by examining how it influences
    $k_{\rm c}$ (the critical longitudinal wavenumber that separates leaky from trapped modes)
    as well as $P$ and $\tau$ for a number of $L/R$.
We found that $k_{\rm c}$ depends only weakly on $\beta_{\rm i}$.
In addition, for both trapped and leaky modes we found that
    $P$ in units of the transverse fast time $R/v_{\rm fi}$ also possesses only a very weak $\beta_{\rm i}$ dependence.
This is attributed to the fact that the effective wavevectors of
    fast sausage modes are largely perpendicular to the background magnetic field.
A weak $\beta_{\rm i}$ dependence of the damping time $\tau$ is also seen, but only when
    $L/R$ exceeds by $\sim 50\%$ some critical value $(L/R)_{\rm c} = \pi/(k_{\rm c} R)$.
Given the sensitive dependence of $k_{\rm c}$ on $\rho_{\rm i}/\rho_{\rm e}$ and $l/R$ as well as
    the specific description of the transverse structuring,
    we conclude that while the much simpler zero-beta theory can be employed for trapped modes,
    effects due to a finite beta should be considered when leaky modes are exploited for seismological purposes.

We note that the dispersion relation (Eq.~\ref{eq_DR}) can find more applications
    than offered here.
First, a finite beta is not specific to flare loops, but exists for hot active region loops
    imaged with, say, SXT~\citep{2007ApJ...656..598W}.
Second, although we examined only fast waves in detail, the DR is equally applicable to
    slow sausage modes in coronal structures for which resonant coupling to torsional Alfv\'en or
    slow waves tends not to appear~\citep[e.g.,][]{2011SSRv..158..289G}.
While one expects slow sausage waves to remain similar to acoustic waves guided by the magnetic field,
    a definitive answer is required to address the effect of a continuous structuring on
    their eigen-frequencies and eigen-functions.
Third, still focusing on fast sausage modes, one may expect that the DR also helps better understand
    the temporal and wavelet signatures of impulsively generated waves in coronal loops with diffuse boundaries,
    as shown in the recent numerical study by \citet{2015ApJ...814..135S}.
The reason is, these signatures are known to critically depend on
    the frequency-dependence of the longitudinal group speeds of trapped modes~\citep[e.g.,][]{1984ApJ...279..857R}.

Nonetheless, the present study has a number of limitations.
First, adopting an ideal MHD approach means that such mechanisms as electron heat conduction and ion viscosity
    are not considered.
While these non-ideal mechanisms were shown by \citet{2007AstL...33..706K}
    to be unlikely the cause for the temporal damping in the QPP event
    reported by~\citet{1973SoPh...32..485M},
   their importance needs to be carefully assessed on a case-by-case basis.
Furthermore, they need to be incorporated when slow sausage waves are of interest.
Second, we did not take into account the longitudinal variations in the plasma or magnetic field strength,
    even though these variations
    are unlikely to be significant for flare loops~\citep{2009A&A...494.1119P}.
Third, this study needs to be extended to account for the singularities in the perturbation equations
    when resonance coupling does occur.
{Focusing on straight tubes, we note that while resonant damping of sausage modes
    does not take place
    when the equilibrium magnetic field is aligned with the tube,
    it is possible when magnetic twist exists~\citep[e.g.,][and references therein]{2016ApJ...823...71G}.
    }
This may, in principle, be done by using the method of singular expansion as adopted in the recent study
    by \citet{2013ApJ...777..158S} who addressed the resonance coupling between fast kink waves
    and torsional Alfv\'en waves in pressureless coronal loops with diffuse boundaries.
{An application of this method will be to examine the resonant damping of sausage modes in magnetically twisted tubes
    with boundaries of arbitrary thickness, thereby generalizing the work by \citet{2016ApJ...823...71G}
    where tube boundaries are assumed to be thin.
}
{Last but not the least, assuming a time-independent equilibrium means that the obtained results
    hold only when the timescale at which the equilibrium parameters vary is substantially longer than
    the wave period $P$, which is of the order of ten seconds given that $P$ is a couple of the transverse fast time.
In reality, however, the physical parameters of flare loops may evolve at a timescale comparable with or even shorter than
    this estimated value of $P$.
There is therefore an imperative need to assess how the temporal variation of the equilibrium affects the properties
    of sausage waves.
Technically speaking, this can be done by either resorting to time-dependent numerical computations such as presented in Appendix~\ref{sec_app_ivp}
    or going beyond the lowest-order treatment of the Wentzel-Kramers-Brillouin (WKB) analysis
    (see e.g., Sect 3.1 in \citeauthor{2007ApJ...661.1222L}~\citeyear{2007ApJ...661.1222L} even though the effects of rapid spatial variation
    on Alfv\'en waves were of interest therein).
An analysis along this line of thinking merits a dedicated study but is beyond the scope of the present manuscript though.
    }

\acknowledgments

This work is supported by the 973 program 2012CB825601,
    the National Natural Science Foundation of China (41174154, 41274176, and 41474149),
    and by the Provincial Natural Science Foundation of Shandong via Grant JQ201212.

\bibliographystyle{apj}
\bibliography{finite_beta}

\IfFileExists{\jobname.bbl}{} {\typeout{}
\typeout{****************************************************}
\typeout{****************************************************}
\typeout{** Please run "bibtex \jobname" to obtain} \typeout{**
the bibliography and then re-run LaTeX} \typeout{** twice to fix
the references !}
\typeout{****************************************************}
\typeout{****************************************************}
\typeout{}}


\begin{center}
{\bf APPENDIX}
\end{center}

\appendix
\section{Coefficients in the Expressions for $y_1$ and $y_2$}
\label{sec_app_coef}

\subsection{Coefficients for General Profiles in the Transition Layer}
\label{sec_app_coef_general}
For general prescriptions for the density and temperature profiles in the transition layer
   as given in Eqs.~(\ref{eq_profile_rho_overall}) and (\ref{eq_profile_T_overall}),
   the coefficients $a_n$
   and $b_n$ in $y_1(x)=\sum\limits^\infty_{n=0}a_nx^n$ and $y_2(x)=\sum\limits^\infty_{n=0}b_nx^n$ are given by
\begin{equation}
  \left\{
  \begin{array}{rcl}
  a_0&=&R^2 \\ [0.3cm]
  a_1&=&0
  \end{array}
  \right. {\rm ~~~and~~~}
  \left\{
  \begin{array}{rcl}
  b_0&=&0 \\ [0.3cm]
  b_1&=&R~.
  \end{array}
  \right.
 \end{equation}
From this point onward, let $\chi$ denote either $a$ or $b$, since both obey the same
    recurrence relations.
In particular, one finds that
\begin{equation}
   \chi_2=-\displaystyle\frac{F_1\chi_1+F_0\chi_0}{G}
 \end{equation}
    where
\begin{equation}
\begin{array}{rcl}
    G&=&4\omega^4RC^2_0+2\gamma\omega^4RC_0V_0-4k^2\omega^2RC^3_0-2\gamma k^2\omega^2RC^2_0V_0
    +4\omega^4RC_0V_0+2\gamma\omega^4RV^2_0\\ [0.15cm]
    &&-8k^2\omega^2RC^2_0V_0-4\gamma k^2\omega^2RV^2_0C_0+4k^4RC^3_0V_0+2\gamma k^4RC^2_0V^2_0 , \\ [0.3cm]
    F_1&=&2\omega^4RC_0V_1-4k^2\omega^2RC^2_0V_1+\gamma\omega^4RC_1V_0+2k^4RC^3_0V_1
    -\gamma\omega^4RC_0V_1+2k^2\omega^2RC^2_0C_1\\[0.15cm]
    &&-2\omega^4RC_1V_0+4k^2\omega^2RV_0C_0C_1+\gamma k^2\omega^2RC^2_0V_1-2k^4RC^2_0V_0C_1-2\omega^4C^2_0-\gamma\omega^4C_0V_0\\[0.15cm]
    &&-2\omega^4C_0V_0+2k^2\omega^2C^3_0+\gamma k^2\omega^2C^2_0V_0-\gamma\omega^4V^2_0+4k^2\omega^2C^2_0V_0+2\gamma k^2\omega^2C_0V^2_0\\[0.15cm]
    &&-2k^4C^3_0V_0-\gamma k^4C^2_0V^2_0 , \\[0.3cm]
    F_0&=&2\omega^6RC_0+\gamma\omega^6RV_0-2k^2\omega^4RC_0V_0
    -\gamma k^2\omega^4RV^2_0-4k^2\omega^4RC^2_0-2\gamma k^2\omega^4RC_0V_0\\[0.15cm]
    &&+4k^4\omega^2RC^2_0V_0+2\gamma k^4\omega^2RC_0V^2_0+2k^4\omega^2RC^3_0+\gamma k^4\omega^2RC^2_0V_0-2k^6RC^3_0V_0-\gamma k^6RC^2_0V^2_0 .
\end{array}
\end{equation}
The coefficients $\chi_i$ for $i \ge 3$ are then given by
\begin{equation}
   \chi_i=-\displaystyle\frac{D(k,\omega^2)}{i(i-1)C(k,\omega^2)}
\label{eq_chi_gt3}
\end{equation}
where
\begin{equation}
\begin{aligned}
   C(k,\omega^2)&=2\omega^4RC^2_0+\gamma\omega^4RC_0V_0-2k^2\omega^2RC^3_0
   -\gamma k^2\omega^2RC^2_0V_0+2\omega^4RC_0V_0+\gamma\omega^4RV^2_0\\
   &-4k^2\omega^2RC^2_0V_0-2\gamma k^2\omega^2RV^2_0C_0+2k^4RC^3_0V_0+\gamma k^4RC^2_0V^2_0
\end{aligned}
 \end{equation}
and
\begin{equation}
\begin{array}{rcl}
   &&D(k,\omega^2)=D_1(k,\omega^2)+D_2(k,\omega^2)+D_3(k,\omega^2)\\[0.3cm]
   &&D_1(k,\omega^2)\\
   &&=\omega^4R\sum\limits_{m=0}^{i-3}\sum\limits_{j=0}^{i-2-m}(m+2)(m+1)( C_{i-2-j-m}+V_{i-2-j-m})(2C_j+\gamma V_j)\chi_{m+2}\\
   &&+\omega^4\sum\limits_{m=0}^{i-3}\sum\limits_{j=0}^{i-3-m}(m+2)(m+1)(C_{i-3-j-m}+V_{i-3-j-m})(2C_j+\gamma V_j)\chi_{m+2}\\
   &&-k^2\omega^2R\sum\limits_{m=0}^{i-3}\sum\limits_{j=0}^{i-2-m}\sum\limits_{l=0}^{i-2-j-m}(m+2)(m+1)(C_{i-2-l-j-m}+2V_{i-2-l-j-m})(2C_j+\gamma V_j)C_l\chi_{m+2}\\
   &&-k^2\omega^2\sum\limits_{m=0}^{i-3}\sum\limits_{j=0}^{i-3-m}\sum\limits_{l=0}^{i-3-j-m}(m+2)(m+1)(C_{i-3-l-j-m}+2V_{i-3-l-j-m})(2C_j+\gamma V_j)C_l\chi_{m+2}\\
   &&+k^4R\sum\limits_{m=0}^{i-3}\sum\limits_{j=0}^{i-2-m}
   \sum\limits_{l=0}^{i-2-j-m}\sum\limits_{s=0}^{i-2-j-l-m}(m+2)(m+1)C_{i-2-j-l-m-s}V_s C_j(2C_l+\gamma V_l)\chi_{m+2}\\
   &&+k^4\sum\limits_{m=0}^{i-3}\sum\limits_{j=0}^{i-3-m}
   \sum\limits_{l=0}^{i-3-j-m}\sum\limits_{s=0}^{i-3-j-l-m}(m+2)(m+1)C_{i-3-j-l-m-s}V_s C_j(2C_l+\gamma V_l)\chi_{m+2}
   \end{array}
\end{equation}
\begin{equation}
\begin{array}{rcl}
&&D_2(k,\omega^2)\\
&&=\omega^4\sum\limits_{m=0}^{i-2}\sum\limits_{j=0}^{i-2-m}[(j+1)(2-\gamma)(C_{i-2-j-m}V_{j+1}-C_{j+1}V_{i-2-j-m})R-2C_{i-2-j-m}
(C_j+V_j)\\ [0.3cm]
&&~~~~~~~~~~~~~~~~~~~-\gamma(C_{i-2-j-m}+V_{i-2-j-m})V_j](m+1)\chi_{m+1}\\[0.3cm]
&&+\omega^4\sum\limits_{m=0}^{i-3}\sum\limits_{j=0}^{i-3-m}(j+1)(2-\gamma)(C_{i-3-j-m}V_{j+1}-V_{i-3-j-m}C_{j+1})(m+1)\chi_{m+1}\\
&&+k^2\omega^2\sum\limits_{m=0}^{i-2}\sum\limits_{j=0}^{i-2-m}\sum\limits_{l=0}^{i-2-j-m}(m+1)[2C_jC_l+(\gamma+4)V_jC_l+2\gamma V_jV_l\\[0.3cm]
&&~~~~~~~~~~~~~~~~~~~~~~~~~~~~~~+(j+1)(4V_lC_{j+1}+\gamma C_lV_{j+1}+2C_lC_{j+1}-4C_lV_{j+1})R]C_{i-2-j-l-m}\chi_{m+1}\\[0.3cm]
&&+k^2\omega^2\sum\limits_{m=0}^{i-3}\sum\limits_{j=0}^{i-3-m}\sum\limits_{l=0}^{i-3-j-m}(j+1)(m+1)C_{i-3-j-l-m}(4V_lC_{j+1}+\gamma C_lV_{j+1}+2C_lC_{j+1}-4C_lV_{j+1})\chi_{m+1}\\
&&+k^4\sum\limits_{m=0}^{i-2}\sum\limits_{j=0}^{i-2-m}
\sum\limits_{l=0}^{i-2-j-m}\sum\limits_{s=0}^{i-2-j-l-m}[2(j+1)(C_sV_{j+1}-V_sC_{j+1})R\\[0.3cm]
&&~~~~~~~~~~~~~~~~~~~~~~~~~~~~~~~~~~~~~~~~-(2C_s+\gamma V_s)V_j]C_{i-2-j-l-m-s}C_l(m+1)\chi_{m+1}\\[0.3cm]
&&+2k^4\sum\limits_{m=0}^{i-3}\sum\limits_{j=0}^{i-3-m}
\sum\limits_{l=0}^{i-3-j-m}\sum\limits_{s=0}^{i-3-j-l-m}(j+1)(m+1)C_{i-3-j-l-m-s}C_l(C_sV_{j+1}-V_sC_{j+1}) \chi_{m+1}
\end{array}
\end{equation}
\begin{equation}
\begin{array}{rcl}
&&D_3(k,\omega^2)\\
&&=\omega^6R\sum\limits_{m=0}^{i-2}(2C_{i-2-m}+\gamma V_{i-2-m})\chi_m+\omega^6\sum\limits_{m=0}^{i-3}(2C_{i-3-m}+\gamma V_{i-3-m})\chi_m\\
&&-k^2\omega^4R\sum\limits_{m=0}^{i-2}\sum\limits_{j=0}^{i-2-m}[2(1+\gamma)C_{i-2-j-m}V_j+\gamma V_{i-2-j-m}V_j+4C_{i-2-j-m}C_j]\chi_m\\
&&-k^2\omega^4\sum\limits_{m=0}^{i-3}\sum\limits_{j=0}^{i-3-m}[2(1+\gamma)C_{i-3-j-m}V_j+\gamma V_{i-3-j-m}V_j+4C_{i-3-j-m}C_j]\chi_m\\
&&+k^4\omega^2R\sum\limits_{m=0}^{i-2}\sum\limits_{j=0}^{i-2-m}\sum\limits_{l=0}^{i-2-j-m}C_{i-2-j-l-m}[(4+\gamma)V_jC_l+2\gamma V_jV_l+2C_jC_l]\chi_m\\
&&+k^4\omega^2\sum\limits_{m=0}^{i-3}\sum\limits_{j=0}^{i-3-m}\sum\limits_{l=0}^{i-3-j-m}C_{i-3-j-l-m}[(4+\gamma)V_jC_l+2\gamma V_jV_l+2C_jC_l]\chi_m\\
&&-k^6R\sum\limits_{m=0}^{i-2}\sum\limits_{j=0}^{i-2-m}
\sum\limits_{l=0}^{i-2-j-m}\sum\limits_{s=0}^{i-2-j-l-m}C_{i-2-j-l-m-s}C_s(2C_j+\gamma V_j)V_l\chi_m\\
&&-k^6\sum\limits_{m=0}^{i-3}\sum\limits_{j=0}^{i-3-m}
\sum\limits_{l=0}^{i-3-j-m}\sum\limits_{s=0}^{i-3-j-l-m}C_{i-3-j-l-m-s}C_s(2C_j+\gamma V_j)V_l\chi_m .
\end{array}
\end{equation}

\subsection{Simplified Coefficients for Profiles Specified in Eq.~(\ref{eq_TL_profile})}
\label{sec_app_coef_para}
When evaluating the coefficients $a_{i}$ and $b_i$ ($i \ge 3$), one finds that
     4-fold summations are necessary.
This turns out to be the most time-consuming part when we numerically solve the dispersion relation.
In fact, it is possible to avoid this because the coefficients
     $C_i~(i>2)$ are all zero for temperature distributions described by
     the profiles chosen in Eq.~(\ref{eq_TL_profile}).
After some algebra, we find that for $i\ge 7$
    the terms $D_1$, $D_2$ and $D_3$ in Eq.~(\ref{eq_chi_gt3}) can be reformulated
    such that only 2-fold summations are involved.
To be specific, they read
\begin{equation}
\begin{aligned}
& D_1(k,\omega^2) \\
&=\sum_{m=0}^{i-3}\sum_{j=0}^{i-2-m}(m+2)(m+1)\chi_{m+2}(2C_j+\gamma V_j)\\
&~~~~~~~~~~~~~~~~\left[\omega^4R(C_{i-2-j-m}+V_{i-2-j-m})-k^2\omega^2RC_0(C_{i-2-j-m}+2V_{i-2-j-m})+k^4RC^2_0V_{i-2-j-m}\right]\\
&+\sum_{m=0}^{i-3}\sum_{j=0}^{i-3-m}(m+2)(m+1)(2C_j+\gamma V_j)\chi_{m+2}[\omega^4(C_{i-3-j-m}+V_{i-3-j-m})\\
&~~~~~~~~~~~~~~~~-k^2\omega^2(RC_1+C_0)(C_{i-3-j-m}+2V_{i-3-j-m})+k^4(2RC_0C_1+C^2_0)V_{i-3-j-m}]\\
&+\sum_{m=0}^{i-4}\sum_{j=0}^{i-4-m}(m+2)(m+1)(2C_j+\gamma V_j)\chi_{m+2}\\
&~~~~~~~~~~~~~~~~\big\{\left[k^4R(C^2_1+2C_0C_2)+2k^4C_0C_1\right]V_{i-4-j-m}-k^2\omega^2(C_1+RC_2)(C_{i-4-j-m}+2V_{i-4-j-m})\big\}\\
&+\sum_{m=0}^{i-5}\sum_{j=0}^{i-5-m}(m+2)(m+1)(2C_j+\gamma V_j)\chi_{m+2}\\
&~~~~~~~~~~~~~~~~\left\{V_{i-5-j-m}\left[2k^4RC_1C_2+k^4(C^2_1+2C_0C_2)\right]-C_2k^2\omega^2(C_{i-5-j-m}+2V_{i-5-j-m})\right\}\\
&+(k^4RC^2_2+2k^4C_2C_1)\sum_{m=0}^{i-6}\sum_{j=0}^{i-6-m}V_{i-6-j-m}(m+2)(m+1)(2C_j+\gamma V_j)\chi_{m+2}\\
&+k^4C^2_2\sum_{m=0}^{i-7}\sum_{j=0}^{i-7-m}V_{i-7-j-m}(m+2)(m+1)(2C_j+\gamma V_j)\chi_{m+2} ,
\end{aligned}
\end{equation}
\newpage
\begin{equation}
\begin{aligned}
&D_2(k,\omega^2)  \\
&=\sum_{m=0}^{i-2}\sum_{j=0}^{i-2-m}(m+1)\chi_{m+1}\big\{\omega^4[(j+1)(2-\gamma)(C_{i-2-j-m}V_{j+1}-C_{j+1}V_{i-2-j-m})R\\
&~~~~~~~~~~~~~~~~-2C_{i-2-j-m}(C_j+V_j)-\gamma(C_{i-2-j-m}+V_{i-2-j-m})V_j]\\
&~~~~~~~~~~~~~~~~+k^2\omega^2C_0[2C_jC_{i-2-j-m}+(\gamma+4)V_jC_{i-2-j-m}+2\gamma V_jV_{i-2-j-m}\\
&~~~~~~~~~~~~~~~~+(j+1)(4V_{i-2-j-m}C_{j+1}+\gamma C_{i-2-j-m}V_{j+1}+2C_{i-2-j-m}C_{j+1}-4C_{i-2-j-m}V_{j+1})R]\\
&~~~~~~~~~~~~~~~~+k^4C^2_0[2(j+1)(C_{i-2-j-m}V_{j+1}-V_{i-2-j-m}C_{j+1})R
-(2C_j+\gamma V_j)V_{i-2-j-m}]\big\}\\
&+\sum_{m=0}^{i-3}\sum_{j=0}^{i-3-m}(m+1)\chi_{m+1}\big\{\omega^4(j+1)(2-\gamma)(C_{i-3-j-m}V_{j+1}-V_{i-3-j-m}C_{j+1})\\
&~~~~~~~~~~~~~~~~+2k^4C^2_0(j+1)(C_{i-3-j-m}V_{j+1}-C_{j+1}V_{i-3-j-m})\\
&~~~~~~~~~~~~~~~~+k^2\omega^2C_1[2C_jC_{i-3-j-m}+(\gamma+4)V_jC_{i-3-j-m}+2\gamma V_jV_{i-3-j-m}\\
&~~~~~~~~~~~~~~~~+(j+1)(4V_{i-3-j-m}C_{j+1}+\gamma C_{i-3-j-m}V_{j+1}+2C_{i-3-j-m}C_{j+1}-4C_{i-3-j-m}V_{j+1})R]\\
&~~~~~~~~~~~~~~~~+k^2\omega^2C_0(j+1)(4V_{i-3-j-m}C_{j+1}+\gamma C_{i-3-j-m}V_{j+1}+2C_{i-3-j-m}C_{j+1}-4C_{i-3-j-m}V_{j+1})\\
&~~~~~~~~~~~~~~~~+2k^4C_0C_1[2(j+1)(C_{i-3-j-m}V_{j+1}-V_{i-3-j-m}C_{j+1})R
-(2C_j+\gamma V_j)V_{i-3-j-m}]\big\}\\
&+\sum_{m=0}^{i-4}\sum_{j=0}^{i-4-m}(m+1)\chi_{m+1}\big\{k^2\omega^2C_2[2C_jC_{i-4-j-m}+(\gamma+4)V_jC_{i-4-j-m}+2\gamma V_jV_{i-4-j-m}\\
&~~~~~~~~~~~~~~~~+(j+1)(4V_{i-4-j-m}C_{j+1}+\gamma C_{i-4-j-m}V_{j+1}+2C_{i-4-j-m}C_{j+1}-4C_{i-4-j-m}V_{j+1})R]\\
&~~~~~~~~~~~~~~~~+k^2\omega^2C_1(j+1)(4V_{i-4-j-m}C_{j+1}+\gamma C_{i-4-j-m}V_{j+1}+2C_{i-4-j-m}C_{j+1}-4C_{i-4-j-m}V_{j+1})\\
&~~~~~~~~~~~~~~~~+4k^4C_0C_1(j+1)(C_{i-4-j-m}V_{j+1}-V_{i-4-j-m}C_{j+1})\\
&~~~~~~~~~~~~~~~~+k^4(C^2_1+2C_0C_2)\left[2(j+1)(C_{i-4-j-m}V_{j+1}-V_{i-4-j-m}C_{j+1})R
-(2C_j+\gamma V_j)V_{i-4-j-m}\right]\big\}\\
&+\sum_{m=0}^{i-5}\sum_{j=0}^{i-5-m}(m+1)\chi_{m+1}\big\{k^2\omega^2C_2(j+1)(4V_{i-5-j-m}C_{j+1}+\gamma C_{i-5-j-m}V_{j+1}\\
&~~~~~~~~~~~~~~~~+2C_{i-5-j-m}C_{j+1}-4C_{i-5-j-m}V_{j+1})\\
&~~~~~~~~~~~~~~~~+k^4\left[4C_2C_1R+2(C^2_1+2C_0C_2)\right](j+1)(C_{i-5-j-m}V_{j+1}-V_{i-5-j-m}C_{j+1})\\
&~~~~~~~~~~~~~~~~-2k^4C_2C_1(2C_j+\gamma V_j)V_{i-5-j-m}\big\}\\
&+\sum_{m=0}^{i-6}\sum_{j=0}^{i-6-m}(m+1)\chi_{m+1}[2(j+1)(k^4C^2_2R+2k^4C_2C_1)(C_{i-6-j-m}V_{j+1}-V_{i-6-j-m}C_{j+1})\\
&~~~~~~~~~~~~~~~~-k^4C^2_2(2C_j+\gamma V_j)V_{i-6-j-m}]\\
&+2k^4C^2_2\sum_{m=0}^{i-7}\sum_{j=0}^{i-7-m}(j+1)(C_{i-7-j-m}V_{j+1}-V_{i-7-j-m}C_{j+1})(m+1)\chi_{m+1} ,
\end{aligned}
\end{equation}
and
\begin{equation}
\begin{aligned}
&D_3(k,\omega^2)	\\
&=\omega^6R\sum_{m=0}^{i-2}(2C_{i-2-m}+\gamma V_{i-2-m})\chi_m+\omega^6\sum_{m=0}^{i-3}(2C_{i-3-m}+\gamma V_{i-3-m})\chi_m\\
&+R\sum_{m=0}^{i-2}\sum_{j=0}^{i-2-m}\chi_m \big\{-k^2\omega^4\left[2(1+\gamma)C_{i-2-j-m}V_j+\gamma V_{i-2-j-m}V_j+4C_{i-2-j-m}C_j\right]\\
&~~~~~~~~~~~~~~~~~~~+k^4\omega^2C_0\left[(4+\gamma)V_jC_{i-2-j-m}+2\gamma V_jV_{i-2-j-m}+2C_jC_{i-2-j-m}\right]\\
&~~~~~~~~~~~~~~~~~~~-k^6C^2_0(2C_j+\gamma V_j)V_{i-2-j-m}\big\}\\
&+\sum_{m=0}^{i-3}\sum_{j=0}^{i-3-m}\chi_m\big\{-k^2\omega^4[2(1+\gamma)C_{i-3-j-m}V_j+\gamma V_{i-3-j-m}V_j+4C_{i-3-j-m}C_j]\\
&~~~~~~~~~~~~~~~~+k^4\omega^2(RC_1+C_0)[(4+\gamma)V_jC_{i-3-j-m}+2\gamma V_jV_{i-3-j-m}+2C_jC_{i-3-j-m}]\\
&~~~~~~~~~~~~~~~~-k^6(2RC_0C_1+C^2_0)(2C_j+\gamma V_j)V_{i-3-j-m}\big\}\\
&+\sum_{m=0}^{i-4}\sum_{j=0}^{i-4-m}\chi_m\big\{k^4\omega^2(RC_2+C_1)\left[(4+\gamma)V_jC_{i-4-j-m}+2\gamma V_jV_{i-4-j-m}+2C_jC_{i-4-j-m}\right]\\
&~~~~~~~~~~~~~~~~-k^6\left[R(C^2_1+2C_0C_2)+2C_0C_1\right](2C_j+\gamma V_j)V_{i-4-j-m}\big\}\\
&+\sum_{m=0}^{i-5}\sum_{j=0}^{i-5-m}\chi_m\big\{k^4\omega^2C_2\left[(4+\gamma)V_jC_{i-5-j-m}+2\gamma V_jV_{i-5-j-m}+2C_jC_{i-5-j-m}\right]\\
&~~~~~~~~~~~~~~~~-k^6\left[2RC_1C_2+(C^2_1+2C_0C_2)\right](2C_j+\gamma V_j)V_{i-5-j-m}\big\}\\
&-k^6(RC^2_2+2C_1C_2)\sum_{m=0}^{i-6}\sum_{j=0}^{i-6-m}(2C_j+\gamma V_j)V_{i-6-j-m}\chi_m\\
&-k^6C^2_2\sum_{m=0}^{i-7}\sum_{j=0}^{i-7-m}(2C_j+\gamma V_j)V_{i-7-j-m}\chi_m .
\end{aligned}
\end{equation}

\section{Standing Fast Sausage Modes in Nonuniform Tubes: An Initial-Value-Problem Approach}
\label{sec_app_ivp}
This section provides an examination from the initial-value-problem (IVP) perspective on the dispersive properties of standing fast sausage modes in magnetic tubes
    for which the transverse density and temperature profiles have been examined in the text.
This is done by directly solving the ideal MHD equations to examine the response of magnetic tubes to an initial transverse velocity perturbation.
We note that a similar study on sausage modes in magnetic slabs with finite gas pressure was carried out by \citet{2009A&A...503..569I}, even though
    different choices for the transverse density and temperature distributions were adopted.
We further note that this practice seems necessary for validating the numerical results presented in the text, because it is independent
    from the eigenmode analysis employed therein.

In view of applications to sausage modes, we solve the axisymmetric
    version of the time-dependent, ideal MHD equations with the PLUTO code \citep{2007ApJS..170..228M} in a standard cylindrical coordinate system
    $(r, \theta, z)$, in which $\theta$ is irrelevant given that $\partial/\partial\theta \equiv 0$.
In addition, the $\theta$-components of the magnetic field $\vec{B}$ and plasma velocity $\vec{v}$ are identically zero.
As implemented by the PLUTO code, only three parameters are needed to normalize the equations.
For this purpose, we choose the mean tube radius $R$, the internal Alfv\'en speed $v_{\rm Ai}$,
    and the density at the tube axis as units for the length, velocity, and density, respectively.
To discretize the equations, a uniform grid with $100$ cells is adopted for the $z$-direction to cover the range from $0$ to $L$.
On the other hand, a nonuniform grid covering the range $[0, r_{\rm M} = 200R]$ is employed in the $r$-direction.
To better resolve wave features close to tube axis, we deploy $200$ cells in a uniform manner for $0 \le  r \le 2 R$,
   but use $400$ cells for $r \ge 2R$ where the grid spacing increases consecutively by a constant factor.
We choose a second-order linear interpolation scheme to reconstruct the piecewise approximation to the primitive vector inside each cell,
   compute the numerical fluxes with the HLLD approximate Riemann solver,
   and advance the equations with a second-order Runge-Kutta marching scheme.
Furthermore, we choose the Constrained Transport method to enforce the divergence-free condition of the magnetic field.
We have made sure that no discernible difference arises in the numerical results if we use a finer grid.
On top of that, we have found no discernible difference when experimenting with some other choices for reconstruction, Riemann solver,
   and time-marching.

Our computations start with a static equilibrium where the transverse density and temperature profiles are described by Eqs.~(\ref{eq_profile_rho_overall}), (\ref{eq_profile_T_overall}) and (\ref{eq_TL_profile}).
The equilibrium magnetic field is set up according to the force balance condition (Eq.~\ref{eq_force_balance}).
An initial perturbation is applied to the transverse velocity $v_r$ only,
    \begin{equation}
     v_r(r,z; t=0) = v_{r0}\sin\left(\displaystyle\frac{\pi z}{L}\right)
       \displaystyle\left(\frac{r}{\sigma}\right)\exp\left[-\displaystyle\frac{r^2}{\sigma^2}\right],
    \label{eq_ini_pert}
    \end{equation}
    which ensures the parity of sausage modes by not displacing the loop axis.
Here $\sigma$ characterizes the extent to which the perturbation spans in the $r$-direction.
We choose $\sigma = R$ to ensure that primarily the lowest-order modes are excited.
In addition, $v_{r0}$ is taken to be $0.05 v_{\rm Ai}$ such that
    nonlinear effects are negligible.

Line-tied boundary conditions are specified at the right boundary $r = r_{\rm M}$, which is
    placed sufficiently far such that the signals in the time interval that we analyze are not contaminated by
    the perturbations reflected off this boundary.
At $r=0$, the boundary conditions are
\begin{equation}\label{eq_r=0}
\begin{array}{cc}
 & v_r(r=0,z;t) = 0, \hspace{0.2cm} B_r(r=0,z; t)= 0, \\ [0.4cm]
 & \displaystyle\frac{\partial \rho}{\partial r}(r=0,z;t)
  =\displaystyle\frac{\partial  v_z}{\partial r}(r=0,z;t)
  =\displaystyle\frac{\partial B_z}{\partial r}(r=0,z;t)
  =\displaystyle\frac{\partial p}{\partial r}(r=0,z;t) = 0~.
 \end{array}
\end{equation}
At $z=0$ and $L$, all physical quantities are fixed at their initial values except $v_z$ and $B_r$, for which we adopt
\begin{equation}\label{eq_z=0L}
  \displaystyle\frac{\partial v_z}{\partial z}(r,z;t)|_{z=0, L}
 =\displaystyle\frac{\partial B_r}{\partial z}(r,z;t)|_{z=0, L}=0~.
\end{equation}

\begin{figure}
\centering
\includegraphics[width=0.9\columnwidth]{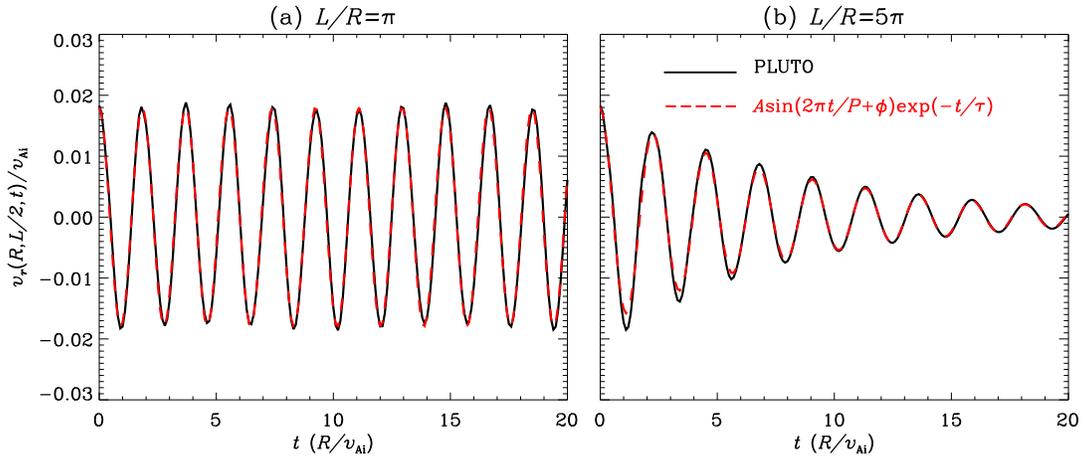}
 \caption{
 Temporal evolution of the transverse velocity $v_r$ associated with sausage perturbations sampled at $[r, z] = [R, L/2]$
     for (a) $L/R = \pi$ and (b) $L/R = 5\pi$.
 For illustration purposes, here we choose a linear profile with $\rho_{\rm i}/\rho_{\rm e} = 30$,
     $l = R$, $\beta_{\rm i} = 0.5$, and $\beta_{\rm e} = 0.01$.
 In addition to the numerical results from the time-dependent computations with the PLUTO code (the black curves),
     the fitting to the curves in the form $A\sin(2\pi t/P+\phi)\exp(-t/\tau)$ is
     given by the red lines for comparison.
 Note that in panel (a), the signal evolves into a sinusoidal form with constant amplitude.
  }
\label{fig_IVP}
\end{figure}

Figure~\ref{fig_IVP} displays the temporal evolution of the transverse velocity $v_r$ sampled at $(r=R, z=L/2)$ for linear profiles
     with $[l/R, \rho_{\rm i}/\rho_{\rm e}, \beta_{\rm i}, \beta_{\rm e}] = [1, 30, 0.5, 0.01]$.
Two values, $\pi$ (Fig.~\ref{fig_IVP}a) and $5\pi$ (Fig.~\ref{fig_IVP}b), are adopted for the length-to-radius ratio to illustrate what happens when
     the mode is in the trapped and leaky regimes, respectively.
In addition to the numerical solutions obtained by PLUTO (the black solid curves), a fitting in the form $A \sin(2\pi t/P+\phi) \exp(-t/\tau)$
     is also shown (the red dashed curves).
From Fig.~\ref{fig_IVP}a one sees that when $L/R = \pi$, the transverse velocity $v_r$ evolves as a sinusoidal signal with a constant amplitude, yielding
     a $P$ of $1.85 R/v_{\rm Ai}$ and $\tau = \infty$.
For comparison, one expects from the eigenmode analysis as presented in Fig.~\ref{fig_Ptau_vs_L} that
     the pertinent eigenmode corresponds to a combination $[P, \tau]$ of $[1.85, \infty] R/v_{\rm Ai}$.
These two values agree closely with each other.
Moving on to Fig.~\ref{fig_IVP}b, one sees that the sampled $v_r$ is well fitted with a decaying sinusoidal signal with $[P, \tau] = [2.27, 8.5] R/v_{\rm Ai}$,
     which is very close to what is found with the eigenmode analysis, namely $[P, \tau] = [2.28, 8.22] R/v_{\rm Ai}$.

Experimenting with a substantial set of profile choices together with different choices
     for the parameters $L/R$, $l/R$, $\rho_{\rm i}/\rho_{\rm e}$, $\beta_{\rm i}$ and $\beta_{\rm e}$,
     we find that the values of $P$ and $\tau$ derived from the time-dependent computations always
     agree very well with those from the eigenmode analysis.
Some examples were shown in Fig.~\ref{fig_Ptau_vs_L} where $P$ and $\tau$ from the fitting procedure were shown by the open circles, whereas
     the results from the eigenmode analysis were shown by the solid curves.
We note that numerical difficulties occur with the PLUTO code when $l/R \ll 1$, in which case the transverse distributions of the density, temperature, and magnetic field
     are all nearly discontinuous around $r=R$.
Despite this minor nuisance, we conclude that standing sausage modes of the lowest-order can be readily excited with our choice of the initial perturbation,
     and their temporal evolution is in close agreement with the expectation from the eigenmode analysis.

\section{A Comparison Between the Present Work and Paper I in the Cold MHD Limit}
A study validating the dispersion relation (DR, Eq.~\ref{eq_DR}) seems necessary, given the complexity in the coefficients intrinsic to the series-expansion-based approach.
This has been partially done with the initial-value-problem approach presented in Appendix~\ref{sec_app_ivp}.
However, it will also be ideal that this DR can be analytically shown to recover some known results in the literature.
In Sect.~\ref{sec_sub_DRtophat} we have shown that when the width of the transition layer ($l$) approaches zero, this DR recovers the
    much-studied result for top-hat profiles.
One may now question whether it is possible to recover the DR derived in cold MHD (Eq.~17 in paper I) by letting the plasma $\beta$ approach zero.
While this is expected given that the two studies differ only in whether a finite plasma $\beta$ is considered,
    we find that this expectation cannot be shown analytically for the time being.
In paper I we started with expanding the density distribution $\rho(r)$ in the transition layer (TL), the coefficients $\rho_n$ were then carried to the coefficients in the expansion
    of the Lagrangian displacement (Eq. 11 in paper I).
In the present work, however, $\rho_n$ does not explicitly appear in the coefficients expressing the perturbation in the TL.
Instead, only the coefficients $V_n$ and $C_n$ in the expansions of the squares of the Alfv\'en and sound speeds appear.
While Eq.~(\ref{eq_coef_Vn}) relating $V_n$ to $\rho_n$ is somehow simplified due to the absence of $C_n$ in this cold MHD limit,
    we find it not straightforward to simplify the coefficients as given in Appendix~\ref{sec_app_coef_general} because multi-fold summations are involved.

This section provides an alternative way to show that the present study yields results that are indeed consistent with paper I when $\beta =0$.
First of all, we note that the time-dependent numerical simulations as presented in Appendix~\ref{sec_app_ivp} have independently verified our finite-$\beta$ DR (Eq.~\ref{eq_DR})
    derived with the eigenmode analysis.
In fact, the pertinent DR in the cold MHD limit (Eq.~17 in paper I) was also validated in the same manner (see Sect.~2.3 in paper I).
It then follows that the two DRs are consistent with each other in the cold MHD limit if they yield identical results.
To this end, we solve Eq.~(\ref{eq_DR}) for an extensive set of profile choices and combinations $[L/R, l/R, \rho_{\rm i}/\rho_{\rm e}]$ with $\beta_{\rm i}$ and $\beta_{\rm e}$ fixed at zero,
   and compare the eigen-frequencies together with eigen-functions with what is found by solving Eq.~(17) in paper I with the same set of parameters.
This comparison shows that both DRs yield identical results.

\begin{figure}
\centering
\includegraphics[width=0.9\columnwidth]{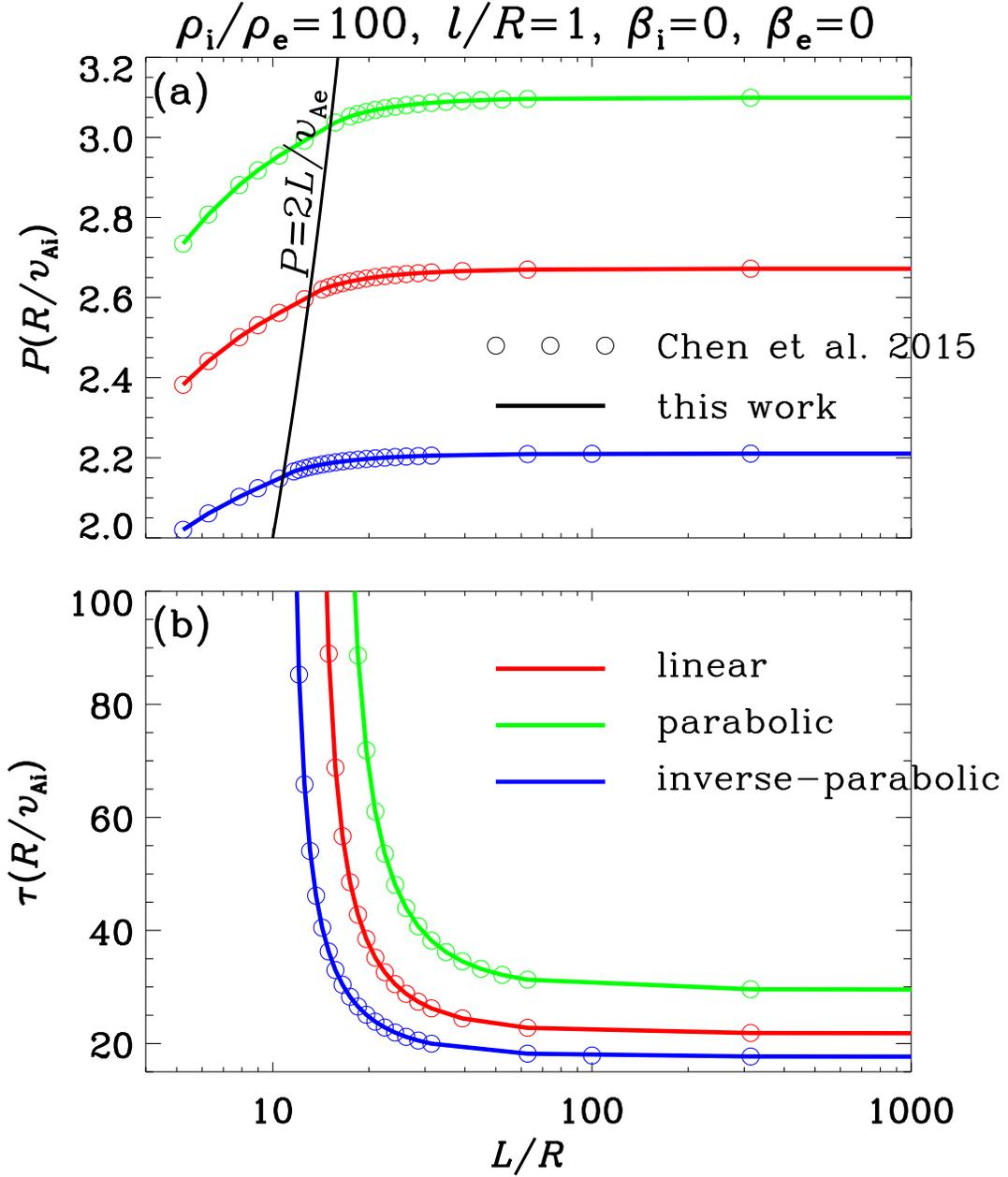}
 \caption{
 Comparison of (a) the periods $P$ and (b) the damping times $\tau$ obtained in the present study (the solid lines)
     and those with the approach in \citet{2015ApJ...812...22C} (open circles)
     for a number of profiles as labeled.
 The solid curves are found by solving Eq.~(\ref{eq_DR}) where we let $\beta_{\rm i}=\beta_{\rm e} =0$,
     while the open circles are found by directly solving the cold MHD dispersion relation
     (Eq.~17 in \citeauthor{2015ApJ...812...22C}~\citeyear{2015ApJ...812...22C}).
 The black solid line in (a) represents $P=2L/v_{\rm Ae}$, which separates the trapped (to its left)
     from leaky (right) modes.
  }
\label{fig_coldcompare_Ptau}
\end{figure}

\begin{figure}
\centering
\includegraphics[width=0.9\columnwidth]{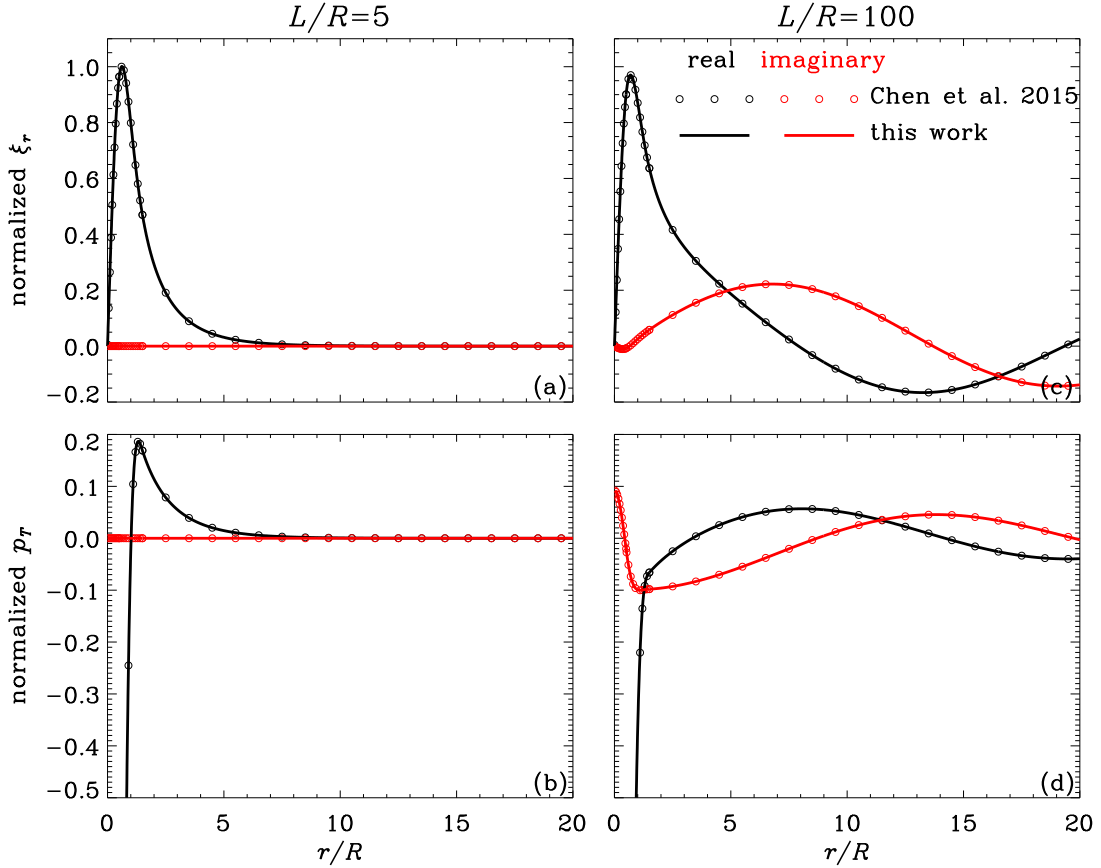}
 \caption{
 Comparison of the Lagrangian displacements (the upper row) and the Eulerian perturbations of total pressure (lower)
     obtained in the present study (the solid lines)
     and those with the approach in \citet{2015ApJ...812...22C} (open circles)
     for an inverse-parabolic profile as labeled.
 Here the left (right) column corresponds to $L/R = 5$ ($100$).
 The solid curves are found with Eqs.~(\ref{eq_y_solution_entire}) and (\ref{eq_Fourie_ptot_xi}) where we let $\beta_{\rm i}=\beta_{\rm e} =0$,
     while the open circles are found with Eqs.~(12) and (13) in \citet{2015ApJ...812...22C}.
 These eigen-functions are normalized such that the magnitude of the Lagrangian displacement attains a maximum of unity.
 Furthermore, the curves and symbols in black (red) represent the real (imaginary) part.
 Note that the sausage mode is trapped in the left column, thereby corresponding to an imaginary part being identically zero.
  }
\label{fig_coldcompare_eigenfunc}
\end{figure}

As an example of this comparison, Fig.~\ref{fig_coldcompare_Ptau} displays the dependence on the length-to-radius ratio $L/R$ of (a) the periods $P$
    and (b) damping times $\tau$ for a number of profiles as labeled.
For illustration purposes, here $[\rho_{\rm i}/\rho_{\rm e}, l/R]$ is chosen to be $[100, 1]$, consistent with Fig.~2 in paper I.
The solid curves represent the results we find by solving Eq.~(\ref{eq_DR}) in which we set $\beta_{\rm i}$ and $\beta_{\rm e}$ to zero,
    whereas the open circles represent what we find by solving the cold MHD DR (Eq.~17 in paper I).
One sees that the values for $P$ and $\tau$ found from both approaches agree with each other exactly.
Figure~\ref{fig_coldcompare_eigenfunc} then displays the spatial distributions of the Lagrangian displacement (the upper row)
    and the Eulerian perturbation of total pressure (lower).
Here an inverse-parabolic profile with $[\rho_{\rm i}/\rho_{\rm e}, l/R]=[100, 1]$ is chosen.
Two values, $5$ (the left column) and $100$ (right) are adopted for the length-to-radius ratio $L/R$.
The curves and symbols in black (red) represent the real (imaginary) part of the eigenfunctions, which are normalized such that
    the magnitude of the Lagrangian displacement attains a maximum of unity.
One sees once again that the approaches presented in the text and in paper I yield identical results in the cold MHD limit.

\clearpage

\clearpage

\clearpage

\clearpage

\clearpage

\clearpage

\clearpage

\clearpage

\clearpage

\clearpage

\end{document}